\documentclass[acmsmall]{acmart}

\usepackage{fancyhdr}
\usepackage[normalem]{ulem}
\usepackage{url}
\usepackage{hyperref}
\usepackage{xcolor}
\usepackage{tabularx}
\usepackage[font={small,bf}]{caption}
\usepackage[font={bf}]{subcaption}
\usepackage{enumitem}
\setlist{nolistsep}
\usepackage{graphicx}
\usepackage{tabu}
\usepackage{multirow}
\usepackage{xspace}
\usepackage[super]{nth}
\usepackage[compact]{titlesec}
\usepackage{textcomp}
\usepackage{bbding}
\usepackage{listings}
\usepackage{fancyvrb}
\usepackage{float}
\newfloat{Listing}{t}{lop}

\newcommand{\Eq}[1]{\hyperref[#1]{Equation~\ref{#1}}}
\newcommand{\Fig}[1]{\hyperref[#1]{Figure~\ref{#1}}}
\newcommand{\Figs}[1]{\hyperref[#1]{Figures~\ref{#1}}}
\newcommand{\Sec}[1]{\hyperref[#1]{Section~\ref{#1}}}
\newcommand{\Secs}[1]{\hyperref[#1]{Sections~\ref{#1}}}
\newcommand{\Tbl}[1]{\hyperref[#1]{Table~\ref{#1}}}
\newcommand{\Lst}[1]{\hyperref[#1]{Listing~\ref{#1}}}

\newcommand{\application}[1]{\textit{#1}}

\AtBeginDocument{%
    \providecommand\BibTeX{{%
        \normalfont B\kern-0.5em{\scshape i\kern-0.25em b}\kern-0.8em\TeX}}}

\setcopyright{acmcopyright}
\copyrightyear{2020}
\acmYear{2020}
\acmDOI{unknown/unknown}

\acmJournal{TOCS}
\acmVolume{**}
\acmNumber{*}
\acmArticle{***}
\acmMonth{7} 

\begin{document}

\title{AI Tax: The Hidden Cost of AI Data Center Applications}

\author{Daniel Richins}
\affiliation{\institution{The University of Texas at Austin}}
\affiliation{\institution{Intel}}
\email{drichins@utexas.edu}

\author{Dharmisha Doshi}
\affiliation{\institution{Intel}}

\author{Matthew Blackmore}
\affiliation{\institution{Intel}}

\author{Aswathy Thulaseedharan Nair}
\affiliation{\institution{Intel}}

\author{Neha Pathapati}
\affiliation{\institution{Intel}}

\author{Ankit Patel}
\affiliation{\institution{Intel}}

\author{Brainard Daguman}
\affiliation{\institution{Intel}}

\author{Daniel Dobrijalowski}
\affiliation{\institution{Intel}}

\author{Ramesh Illikkal}
\affiliation{\institution{Intel}}

\author{Kevin Long}
\affiliation{\institution{Intel}}

\author{David Zimmerman}
\affiliation{\institution{Intel}}

\author{Vijay Janapa Reddi}
\affiliation{\institution{Harvard University}}
\affiliation{\institution{The University of Texas at Austin}}

\renewcommand{\shortauthors}{Richins et al.}

\thanks{This work extends a previous article published in the HPCA 2020 Industry Session: \textit{Missing the Forest for the Trees: End-to-End AI Application Performance in Edge Data Centers}~\cite{richins2020missing}.  This article adds (1) a discussion on the need for end-to-end AI benchmarks; (2) additional data and explanation for the deployment choices made in \application{Face Recognition}; (3) examination of the one of the causes of broker waiting time; and (4) AI tax and acceleration analysis of a second edge data center workload.}

\begin{abstract}

Artificial intelligence and machine learning are experiencing widespread adoption in industry and academia.  This has been driven by rapid advances in the applications and accuracy of AI through increasingly complex algorithms and models; this, in turn, has spurred research into specialized hardware AI accelerators.  Given the rapid pace of advances, it is easy to forget that they are often developed and evaluated in a vacuum without considering the full application environment.  This paper emphasizes the need for a holistic, end-to-end analysis of AI workloads and reveals the ``AI tax.''  We deploy and characterize \application{Face Recognition} in an edge data center.  The application is an AI-centric edge video analytics application built using popular open source infrastructure and ML tools.  Despite using state-of-the-art AI and ML algorithms, the application relies heavily on pre- and post-processing code.  As AI-centric applications benefit from the acceleration promised by accelerators, we find they impose stresses on the hardware and software infrastructure: storage and network bandwidth become major bottlenecks with increasing AI acceleration.  By specializing for AI applications, we show that a purpose-built edge data center can be designed for the stresses of accelerated AI at 15\% lower TCO than one derived from homogeneous servers and infrastructure.  

\end{abstract}



\maketitle

\section{Introduction}
\label{sec:intro}

Artificial intelligence (AI), especially the field of machine learning (ML), is transforming the marketplace.  Sparked by advances in computer system design, enterprises are leveraging AI in every possible manner to provide unprecedented new services to their end users, ranging from recommendation-based online shopping and personalized social network services to virtual personal assistants and better health care.

To enable ML, there has been a flurry of work at two extremes.  At one extreme is the effort that focuses on hardware acceleration of ML kernels~\cite{habana, chen2014diannao, tpu1, intelncs2, nvidiaai}. At the other extreme is the effort that focuses on engineering the system and its supporting infrastructure, such as the associated networking and storage.  The former is essential for enabling microprocessor advancements, while the latter is essential for allowing cloud-scale deployment.

But recent years have seen a shift in the needs of the industry.  While much research has been dedicated to maximizing and accelerating machine learning performance, recent industry perspectives have urged for a more holistic understanding of machine learning development and performance.  Facebook, for example, has discussed some of the challenges it has faced running AI at scale and encouraged research on mitigating those challenges~\cite{hazelwood2018applied}.
Instead of focusing solely on the AI kernel computation time, there is a need to look at the bigger picture.  Enabling AI applications involves several stages: ingesting the data, pre-processing the data, offloading the data to an AI accelerator, waiting for data, post-processing the result, etc., all of which affect the requests' end-to-end latency and total system throughput.

At the same time, the industry is witnessing AI services migrate from warehouse-scale systems to smaller purpose-built data centers located at the edge, closer to end-users~\cite{hpe2018onprem}. These edge data centers complement existing cloud- or large-scale services by being physically closer to the data source, which enables faster responses to latency-sensitive or bandwidth-hungry application services~\cite{vxchnge-edgelatency}.  There are also data sovereignty and regulatory compliance rules to safeguard data privacy that are addressed with edge data centers~\cite{cloud-sovereignty}.  Moreover, many mid-size organizations find it more economical to invest in on-premise data centers that are purpose-built for executing a particular type of task~\cite{data-economy}. So despite the continued growth in public cloud solutions, spending for edge data centers is predicted to increase~\cite{emconit}.

In this work, we study the intersection of user-facing AI computing---the inference side of machine learning---and smaller, edge data centers to reveal the often overlooked ``AI tax'': the additional compute cycles, infrastructure, and latency required to support the AI at the application's heart.
In the context of a data center, execution of a fully developed, deployment-ready AI-centric application relies on more than just AI algorithms.  End users' requests demand pre-processing to ready them for the pipeline; intermediate data must be communicated between stages, often over a network using custom protocols; the communication framework often has built-in data reliability safeguards which impose overheads on data movement; and each stage faces its own overhead for moving data.  All of these components together add to the overhead of executing AI.

We study a full deployment of \application{Face Recognition}, an end-to-end video analytics AI-centric application at the edge.
Our setup is an industry deployment of the Google FaceNet~\cite{schroff2015facenet} architecture in an edge data center.  Our application is entirely focused on the inference side of machine learning, where it is exposed to end-users and faces associated latency constraints.  As an AI-centric application, \application{Face Recognition} is a good choice as it employs three distinct artificial intelligence algorithms, including two neural networks and a classification algorithm.  Furthermore, it is representative of the reality of a considerable portion of AI and ML applications: many AI applications exist as streaming services, deployed in data centers, serving real-time needs of consumers.  Coordinating the many activities required to transform raw data into useful, easily consumable conclusions requires the intricacies and nuances of any distributed application: networking equipment, storage devices, coordination, data durability, power distribution, cooling, communication protocols, data compression, etc.~\cite{kanev2015profiling}.

We find that in today's edge data centers, already the communication framework can constitute over 33\% on the latency of the application.  \application{Face Recognition} is built on top of Apache Kafka~\cite{apachekafka}, which is widely adopted both directly and as a fabric upon which advanced streaming frameworks are built~\cite{datanami, storm-kafka, samza-kafka, flink-kafka, apex-kafka, druid-kafka}.  Kafka is also representative of alternative frameworks that utilize communication hot spots.  The simplicity and impressive performance of Apache Kafka have established it as a common denominator for many industry-quality projects.  Despite this, requests can spend substantial time passing through the framework.

Moreover, we show that as accelerator technologies advance and integrate into production environments, the supporting portions of the pipeline will soon supplant AI as the primary determinant of performance.  We measure the implications of greater AI inference acceleration.  Apache Kafka becomes increasingly stressed to move the vastly increased volume of data ingested by the application.  Even at relatively low acceleration factors, the added stress will quickly overwhelm Kafka's current capabilities.  We demonstrate that at a very modest 8$\times$ acceleration factor, Kafka overwhelms the capabilities of its underlying storage.

These findings present a unique opportunity on the compute research spectrum: rather than neglecting the execution context of AI and without moving into the realm of cloud compute where resources must be generic and homogeneous enough to handle all kinds of workloads, we show a proof-of-concept for the economic value of edge data centers.  We demonstrate how a data center that is custom-built for the needs of a streaming AI workload can accommodate the anticipated requirements of accelerated AI without over-provisioning, thereby realizing an overall decrease in the total cost of ownership (TCO) in excess of 15\% over a homogeneous edge data center.

Although our deep-dive analysis primarily focuses on edge video analytics with \application{Face Recognition} (as the poster child application), we also conduct a basic analysis of a second application, \application{Object Detection}, deployed similarly to \application{Face Recognition}, and show that it faces comparable AI tax challenges.  In other words, our analyses and conclusions are not specific to one application; we further discuss how the underlying infrastructure of an end-to-end AI application will present mainly the same bottlenecks regardless of the AI application.

In summary, our main contributions and insights are

\begin{enumerate}
    \item Where much focus is devoted to tuning and accelerating AI inference to enable faster compute, we instead evaluate the larger \textbf{system-level implications of end-to-end AI applications and expose the AI tax};
    \item We show that the \textbf{general-purpose CPU performance remains a significant determinant of overall request performance} because processing an end-user request requires more than just AI kernel computation;
    \item The \textbf{communication layer of an AI application imposes a large overhead} on the latency of processing;
    \item The \textbf{increased throughput from AI acceleration will overwhelm the communication substrate}; and
    \item A \textbf{purpose-built data center can adapt to the upcoming challenges of accelerated AI at a lower TCO} than a generic, homogeneous data center.
\end{enumerate}

There are, of course, innumerable ways to deploy a data center AI application.  In this research, we focus on a scheme that concentrates communication in a few data brokers.  We expect that our main takeaways will be applicable to any deployment that utilizes some sort of data broker.  Namely, with the increasing data throughput of increasingly effective AI accelerators, the brokers will become a point of failure.  The amount of data moving through the application can quickly overwhelm the capabilities of the brokers.

The proper solutions to the AI tax may reside in a customized edge data center, as we suggest, or they may depend on a different deployment scheme.  In any case, our work clearly demonstrates the criticality of understanding AI applications from an end-to-end viewpoint.  Without that perspective, we would have no comprehension of the AI tax nor any reason to suspect that it would become the limiting factor to performance.

The remainder of this paper is structured as follows.  In \Sec{sec:aiscope}, we motivate study of AI applications in a full, end-to-end, data center context.  In \Sec{sec:application}, we introduce our primary application, \application{Face Recognition}, and explain its deployment and optimization in our edge data center.  In \Sec{sec:aitax}, we elucidate the AI tax, characterizing the performance and limitations of the end-to-end AI application.  In \Sec{sec:accelerate}, we conduct a forward-looking analysis of \application{Face Recognition} under accelerated AI compute and identify significant impediments to improving performance.  In \Sec{sec:discussion}, we show that the AI tax exposition is not unique to \application{Face Recognition} by similarly deploying and evaluating a second application.  In \Sec{sec:solution}, we show that an edge data center can be purpose-built to address the upcoming challenges of AI while reducing TCO.  We then distinguish our work from prior art in \Sec{sec:related} and conclude in \Sec{sec:conclusion}.

\section{AI in Edge Data Centers}
\label{sec:aiscope}

\begin{figure}
    \includegraphics[width=\textwidth]{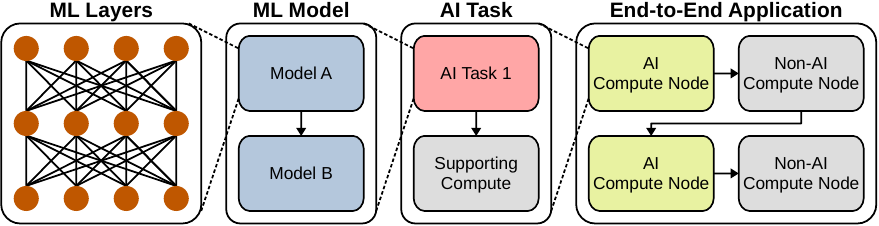}
    \caption{Increasing levels of abstraction in understanding AI applications.  At its most basic level, a machine learning AI application is computation of layers through common operations like matrix multiplication.  Layers are used in combination to create an ML model.  One or more models combine with supporting compute to compose AI tasks.  And finally, AI tasks and additional supporting infrastructure is distributed in a data center to create a complete application.}
    \label{fig:aiscope}
\end{figure}

With the explosive growth of AI applications, much work has been done to try to characterize and understand them through numerous proposed benchmark suites.  Ranging from device-level computations to complete applications, these benchmarking efforts cover a range of abstraction levels, each recognizing the need to understand AI applications from a variety of abstraction levels (\Fig{fig:aiscope}).  In this section, we explore the range of abstractions and explain the need for the next level of abstraction: edge data center, end-to-end AI applications.  In \Sec{sec:aiscope:layers}, we look at benchmarking efforts aimed at understanding ML layers.  \Sec{sec:aiscope:model} looks at complete ML models built from potentially many layers of compute.  \Sec{sec:aiscope:task} considers entire AI tasks composed of potentially multiple ML models operating in tandem.  Finally, \Sec{sec:aiscope:dc} takes the next logical step and motivates the study of a complete AI application as it exists in an industry-quality edge data center.


\subsection{ML Layers}
\label{sec:aiscope:layers}

At its heart, an AI application is a set of computations done on some piece of hardware, and, though this seems simplistic, it is critical to understand what computations compose the application and how they fit together (\Fig{fig:aiscope}, ``ML Layers'').  DeepBench~\cite{deepbench} enables researchers to benchmark the primitive layers of ML models (matrix multiplies, convolutions, and recurrent operations) at the lowest level---on CPU and GPU using primitive machine learning libraries.  It exists below any ML framework (such as TensorFlow~\cite{tensorflow} and Caffe~\cite{caffe}).  Its goal is to elucidate the performance of the most common operations on various hardware devices.


Taken in isolation, though, DeepBench is incomplete.  Basic operations like matrix multiplication are essential in machine learning, but in practice they occur only in certain model layers and within specific data access patterns that can vary from one ML framework to another.  These details, available only in the next level of abstraction, give important context for the basic operations of benchmarks like DeepBench.

\subsection{ML Model}
\label{sec:aiscope:model}

The next level of abstraction is the ML model, built from fundamental operations and layers (\Fig{fig:aiscope}, ``ML Model'').  A complete model will typically operate on its input (such as an image) to produce a useful evaluation (such as identifying the objects in the image).  Benchmarking complete models is critical because it determines precisely which operations are to be performed, how common each operation is, and the computational intensity of operating.


There are numerous benchmark suites that address this level of abstraction.  AI Matrix~\cite{zhang2019ai,AIMatrix} introduces numerous complete ML models, covering such diverse application domains as image classification and neural machine translation, and also provides for synthetic models that are generated to mimic desired workload characteristics.  MLMark, an EEMBC benchmark, is focused on benchmarking ML inference on embedded or edge devices~\cite{torelliMLMark}.  AI-Benchmark~\cite{AIBenchmark} is designed specifically to target smartphone performance~\cite{ignatov2018ai, ignatov2019ai}.  AIIA-Benchmark targets accelerator devices and aims to provide a useful means of comparison.




While these benchmark suites are indispensable, they are also incomplete.  ML models do not exist or execute in a vacuum but rely on supporting compute.  In practice, this often means that, unlike these benchmarks, ML model execution cannot operate uninterrupted but must be supported by additional models or even non-ML compute.

\subsection{AI Task}
\label{sec:aiscope:task}

For ML to be useful, it needs the help of supporting compute, i.e.\ pre- and post-processing.  Additionally, a model is often only one piece of a series of models working as a unit to produce a result.  Together, these compose an AI task (\Fig{fig:aiscope}, ``AI Task'').  For example, before a model can operate on an input image, the image must be transformed into the proper size, with the proper color encoding, in the proper layout, to match the requirements of the model; the model output must similarly undergo transformation to be prepared for the next stage of processing or to be returned to the user.  At this level of abstraction, the real behavior of an AI application starts to become apparent.



Without the pre- and post-processing steps, the AI kernel is basically useless.  Even so, we are aware of no benchmarking suite that captures this more complete, context-aware, AI task-level view.

While an AI task may represent a complete application, if it is executing on a single device, much of the AI compute in practice takes place as real-time services provided by industry players for end users.  In such a scenario, an AI task is itself incomplete, as an AI task will be part of a larger application that spans multiple servers and interacts over the network both internally and with end users.

\subsection{End-to-End Application}
\label{sec:aiscope:dc}

\newcommand{\chk}[0]{{\normalsize \checkmark}}
\newcommand{\qmk}[0]{{\normalsize \bf ?}}
\begin{table}[t]
    \tiny
    \caption{Comparison of various ML benchmarks showing how much of the end-to-end AI application is captured by each.  While most benchmarks thoroughly cover the ML layers and model space, a complete application is far larger than these computational kernels, and none of the existing benchmarks capture the whole picture.}
    \begin{tabularx}{1.0\textwidth}{ p{2cm} p{1.1cm} p{1.1cm} p{1.1cm} p{1.1cm} p{1.1cm} p{1.1cm} p{1.1cm} p{1.1cm} }
        \toprule
        Benchmark           & ML \newline Layers& ML Model          & AI Task           & Orches\-tration Software & Pre-Processing & Post-Processing & Network         & Storage           \\ \midrule
        DeepBench           & \chk              &                   &                   &                   &                   &                   &                   &                   \\
        AI Matrix           & \chk              & \chk              &                   &                   &                   &                   &                   &                   \\
        MLMark              & \chk              & \chk              &                   &                   &                   &                   &                   &                   \\
        AI-Benchmark      & \chk              & \chk              &                   &                   &                   &                   &                   &                   \\
        AIIA-Benchmark    & \chk              & \chk              &                   &                   &                   &                   &                   &                   \\
        DAWN\-Bench         & \chk              & \chk              &                   &                   &                   &                   &                   &                   \\
        MLPerf              & \chk              & \chk              &                   &                   &                   &                   &                   &                   \\ \midrule
        End-to-End      & \qmk              & \qmk              & \qmk              & \qmk              & \qmk              & \qmk              & \qmk              & \qmk              \\
        \bottomrule
    \end{tabularx}
    \label{tbl:benchmarks}
\end{table}

The highest level of abstraction is the data center-level, where we finally see the end-to-end application (\Fig{fig:aiscope}, ``End-to-End Application'').  The various AI tasks are deployed to different nodes throughout the data center along with their supporting compute.  Additionally, some nodes may be entirely dedicated to non-AI, supporting compute.  \textit{Of critical importance, at the data center level, the networking equipment, storage devices, communication protocols, data center management software, and application coordination software all become part of the AI application.}


The numerous past ML benchmarking efforts are invaluable for understanding the heart of AI applications; however, they all fall short of this highest level of abstraction.  \Tbl{tbl:benchmarks} illustrates, for a small sampling of benchmarks, how much of an AI application is \textit{not captured} and \textit{not understood} as a result of failing to rise to this level of abstraction.  All the benchmarks we have mentioned in this section do a good job of benchmarking ML compute details, but they all leave gaping holes in the understanding of end-to-end application performance.
DAWNBench~\cite{DAWNBench} and MLPerf~\cite{mlperf} are noteworthy for trying to introduce greater realism into benchmarking ML models; unfortunately, they do not go far enough.  DAWNBench recognizes the importance of batch sizing~\cite{coleman2017dawnbench}, which is widely known and adopted in realistic deployments of AI applications to maximize performance.  MLPerf takes this further and introduces a number of ``scenarios'' under which ML models can be evaluated.  These scenarios incorporate the essential concept of latency-bounded throughput---maximizing throughput while honoring latency constraints.  They try to mimic a setup that would exist in a real-world data center where AI applications are deployed~\cite{janapareddi2019mlperf}.

Ultimately, however, even though these benchmarks acknowledge that ML models are executed in a server-like scenario, they sidestep the issue.  MLPerf, for example, attempts to mimic a server setup by having requests arrive at random intervals according to a Poisson distribution, but this ignores the portion of the pipeline that actually provides and pre-processes the requests.

AIBench is an ambitious undertaking that recognizes the need for end-to-end benchmarks~\cite{gao2019aibench}.  Designed to be easily extensible by building from a common framework, AIBench has implemented two scenarios: e-commerce searching and online language translation.  Both scenarios operate in a data center, utilize multiple stages of compute and inference coordinated by orchestration software, rely on network and storage, and utilize pre- and post-processing.  However, as it stands, AIBench is insufficient to cover the breadth and depth of AI applications and has not undergone the extensive and holistic evaluation for which we argue in this paper.  For these reasons, we consider it more of a standalone application than a true benchmark.


To fully understand an AI application requires taking a holistic view at this highest level of abstraction.  We are not aware of any effort to capture this complete understanding.  While we do not present a new benchmark suite, in this work we undertake to present a thorough, forward-looking, end-to-end evaluation of a complete AI application.

\section{End-to-End Video Analytics Application}
\label{sec:application}

Given the importance of understanding the end-to-end AI application as deployed in a data center, we describe in this section the AI-centric \application{Face Recognition} application we developed and how we deployed it.  It is based on Google's FaceNet algorithm~\cite{schroff2015facenet}.  However, \application{Face Recognition} is much more than just FaceNet---it is a full data center application.  Going from the algorithm to the full workload requires algorithm partitioning, containerization, work coordination, and communication management.  We explain how the logical flow of \application{Face Recognition} is transformed into a functional streaming data center application.

We chose a video analytics application for our studies of AI tax due to the rising importance of this domain.  The global market for video analytics is expected to hit US\$25 billion by 2026 due to rapid adoption of video technologies across industries such as retail, manufacturing, and smart cities~\cite{marketwatch}.  Video analytics uses AI to provide cost-efficient business intelligence insights to its users.  The domain is slated for deployment in edge data centers, as opposed to traditional cloud- or warehouse-scale systems due to latency constraints, network bandwidth, and privacy regulations.

In \Sec{sec:application:workload} we introduce and describe \application{Face Recognition}.  \Sec{sec:application:setup} summarizes our edge data center setup.  In \Sec{sec:application:stages}, we explore different ways of deploying it to the data center nodes.  In \Sec{sec:application:kafka}, we introduce the Apache Kafka framework, which we use to coordinate the application steps and manage communication between them.  \Sec{sec:application:container} explores the design space for individual containers while considering both latency and throughput constraints.  In \Sec{sec:application:ratios}, we explain how \application{Face Recognition} is deployed at scale.

We detail the deployment of a second data center-level application in \Sec{sec:discussion}.

\begin{figure}[t]
    \centering
    \includegraphics[width=0.7\textwidth]{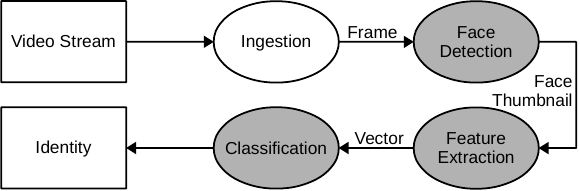}
    \caption{Algorithmic flow of \application{Face Recognition}.  A video stream enters \emph{ingestion} for separation into individual frames.  \emph{Face detection} finds any faces within a frame and produces a thumbnail for each.  \emph{Feature extraction} generates identifying features for each face.  Finally, \emph{classification} finds a nearest match to known faces to produce an identity.  The shaded ovals indicate the stages where AI compute takes place.}
    \label{fig:fr_flow_algorithm}
\end{figure}

\subsection{Video Analytics Pipeline}
\label{sec:application:workload}

Video analytics is the automatic analysis of video data.  For our analysis, we developed and deployed a video analytics application for edge usage called \application{Face Recognition} (\Fig{fig:fr_flow_algorithm}).  Though it uses machine learning, this application is strictly user-facing, i.e.\ it uses inference rather than training.

Our implementation of \application{Face Recognition} relies heavily on artificial intelligence and machine learning algorithms implemented in Tensor\-Flow~\cite{abadi2016tensorflow}.  Given a number of input video streams, the application parses the videos into individual frames, locates faces within the frames, and identifies each face as a certain individual.  The video streams could represent a surveillance system's cameras~\cite{regazzoni2010video}, offline processing of recorded videos, a transaction-less shopping environment~\cite{amazongo}, or many other applications where multiple streams are concurrently being fed into the system.

The \application{Face Recognition} application consists of four primary processing stages (\Fig{fig:fr_flow_algorithm}) and is built from MT-CNNs (multi-task cascaded convolutional networks)~\cite{zhang2016joint} along with Google's Face\-Net~\cite{schroff2015facenet, githubfacenet}.  Like many real-world use cases, the application involves multiple inferences per query.
\begin{enumerate}
    \item \textbf{Ingestion} is a pre-processing stage that ingests a video stream and parses it into individual frames.  This stage is critical as FaceNet cannot operate directly on video.
    \item \textbf{Face detection (AI)} relies on MT-CNNs to detect any faces within a frame without making any effort to identify them.  It determines bounding boxes for and produces a \texttt{160x160} thumbnail of each face in a frame. 
    \item \textbf{Feature extraction (AI)} is built using the Inception-Resnet~\cite{szegedy2017inception} architecture and produces a 128-byte vector of essential features that describe each face. 
    \item \textbf{Classification (AI)} compares the feature vector for a face against a set of known face vectors to find the best match by means of a support vector machine (SVM), yielding an identification. 
\end{enumerate}

\application{Face Detection} is designed as a pipelined streaming application---it ingests video streams at or near their native frame rate, injects them into the pipeline, and yields facial identities for frames after some delay.  Even though the overall latency may exceed the time between adjacent frames in a video stream, because the application is pipelined, the throughput is at or near the native frame rate.

\subsection{Edge Data Center}
\label{sec:application:setup}

\begin{table}[t]
    \small
    \centering
    \caption{Server details.  Each server in our data center is well-equipped, using leading-edge technology.  Our nodes are powered by Intel Xeon Platinum 8176 or comparable CPUs.}
    \begin{tabu}{ l l }
        \toprule
        \textbf{Component} & \textbf{Details} \\ \midrule
        CPU & 2$\times$ Intel Xeon Platinum 8176~\cite{xeon8176} \\
        ~~~~\textit{Cores} & ~~~~\textit{28} \\
        ~~~~\textit{Base Frequency} & ~~~~\textit{2.10~GHz} \\
        ~~~~\textit{Max Turbo Frequency} & ~~~~\textit{3.80~GHz} \\
        ~~~~\textit{SMT} & ~~~~\textit{2-way} \\
        ~~~~\textit{LLC} & ~~~~\textit{38.5~MB} \\
        Memory & 384 GB DDR4-2666~\cite{p4510} \\
        Storage & Intel SSD P4510 \\
        ~~~~\textit{Read BW} & ~~~~\textit{2.85 GB/s} \\
        ~~~~\textit{Write BW} & ~~~~\textit{1.1 GB/s} \\
        ~~~~\textit{Read Latency} & ~~~~\textit{77 us} \\
        ~~~~\textit{Write Latency} & ~~~~\textit{18 us} \\
        Network & Full duplex 100~Gbps Ethernet \\
        \bottomrule
    \end{tabu}
    \label{tbl:dc_details}
\end{table}

For our experiments, we utilize a small edge data center built from high-performance servers (see \Tbl{tbl:dc_details}).  We use over 2200 processor cores spread across 40+ nodes to ensure that we have a realistic deployment whose characteristics can scale to larger setups.  Each node is equipped with 56 physical cores spread across two sockets, 384~GB of RAM, high-speed local NVMe storage, and 100~Gbps Ethernet.  The nodes are connected in a fat tree topology~\cite{leiserson1985fat}.

We rely on industry-standard, open-source tools for our deployment.  After dividing the application algorithm into steps, we deploy the different steps in lightweight Docker~\cite{docker} containers.  Depending on the resource requirements of a container, it may be deployed in isolation on a data center node or it may be deployed alongside other containers on the same node.  The deployment of the various containers is managed using Kubernetes~\cite{kubernetes}.  As the application steps are separated from each other, data has to be communicated between them; for this, we rely on Apache Kafka~\cite{apachekafka}.  We explore these and other details in the remainder of this section.

\subsection{Stage Count}
\label{sec:application:stages}

As an application, \application{Face Recognition} separates its algorithmic steps into discrete stages that coordinate with one another while running independently on separate nodes to produce a legitimate data center application.  Separating the algorithm into multiple coordinated steps allows different stages of the application to adapt to the speed and requirements of other stages.

Even though there are four logical steps in the \application{Face Recognition} algorithm (ingestion, face detection, feature extraction, and classification), in practice this reduces to three.  Feature extraction and classification are tightly coupled and their code is hard to separate.  We term the combined steps, ``identification.''  This simplification results in a reduced design exploration space.

\begin{figure}
    \small
    \centering
    \begin{subfigure}{0.55\textwidth}
        \small
        \centering
        \includegraphics[width=\textwidth]{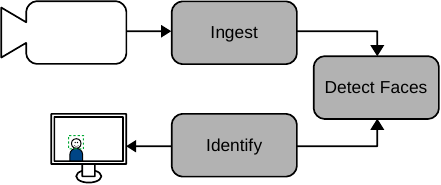}
        \caption{Three-stage pipeline.}
        \label{fig:fr_stagecounts:three}
    \end{subfigure}
    \hspace{1ex}
    \begin{subfigure}{0.41\textwidth}
        \small
        \centering
        \includegraphics[width=\textwidth]{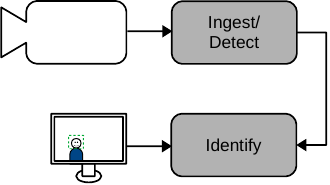}
        \caption{Two-stage pipeline.}
        \label{fig:fr_stagecounts:two}
    \end{subfigure}
    \caption{Two possibilities for deploying \application{Face Recognition} as a data center application.}
    \label{fig:fr_stagecounts}
\end{figure}

In optimizing \application{Face Recognition}, we explored two designs for separating the algorithm into stages, shown in \Fig{fig:fr_stagecounts}.  In \Fig{fig:fr_stagecounts:three}, the application is separated into its three logical components: ingestion, face detection, and identification.  The ingestion stage, like the algorithmic step of the same name, is fed a video stream which is parsed into separated video frames.  Because the ingestion and face detection stages exist in separate containers, likely on physically distinct nodes within the data center, the separated video frames must be transferred between them over the network.  Similarly, the cropped face thumbnails produced by the face detection stage are sent over the network to the identification stage.

The alternative deployment for \application{Face Recognition} is shown in \Fig{fig:fr_stagecounts:two}: ingestion and face detection are combined into a single \textit{ingest/detect} stage which operates along with identification.  In this design, ingestion and face detection processes operate within the same container, so frames are transferred directly between them, leaving only the face thumbnails to be sent over the network to the identification stage.

Beyond the obvious difference between the two- and three-stage designs (the three-stage design imposes greater demands on the network), there is a subtler and more profound difference that must be considered.  In the three-stage design, the junction between the ingestion and face detection stages is very simple: ingestion always produces one frame at a time at a particular rate and each frame must run through face detection exactly once.  In contrast, the junction between face detection and identification transfers a variable number of faces and this face count determines the amount of work to be done in identification.  Hence, the compute demands placed on the identification stage will vary based on the nature of the video streams---a video stream that captures many faces will demand greater identification processing power.  Thus, by decoupling face detection from identification, we create a point of flexibility where a single face detection (or ingest/detect) container can be serviced by potentially many identification containers---i.e.\ load balancing.  The junction between ingestion and face detection has no such requirement.

Both in order to reduce the demands on the network and to combine processing steps where it makes sense to do so, we adopt the two-stage design (\Fig{fig:fr_stagecounts:two}).  We also utilize one additional container, the \textit{broker}, that we discuss in \Sec{sec:application:kafka}.

The \textit{ingest/detect} container runs two processes internally, one for ingestion and the other for face detection.  Ingestion processes a video stream (in our experiments, we use a \texttt{1920x1080} video file for deterministic operation) and parses the stream into frames.  It resizes the frames to \texttt{960x540} before passing them to the face detection process.  Face detection produces a thumbnail for each face in a frame, if any (our video yields zero to five faces and averages 0.64 faces per frame, with face thumbnails averaging 37~kB each).  If no faces are found, identification is not needed; otherwise the faces are transferred to the \textit{identification} container.  It consists of a single process (the combined feature extraction and classification, as mentioned earlier) which processes face thumbnails to yield an identity.

In accordance with industry practice, we execute all inference directly on the CPU~\cite{hazelwood2018applied}.  This yields the lowest latency which is critical in a user-facing application.

\subsection{Apache Kafka}
\label{sec:application:kafka}

Communication between containers running on separate nodes within a data center requires intelligence and elegance.  We rely on Apache Kafka~\cite{apachekafka} to manage the communication between ingest/detect and identification, allowing for load balancing and offering rapid adaptation in the presence node failures.  Though there exists a variety of open source tools for building and managing streaming applications~\cite{apachestorm, apachesamza, apacheflink, apacheapex, apachedruid, kafkastreams}, we note that these tools tend to rely on a separate framework for enabling communication between containers.  It is common in practice to rely on Apache Kafka to serve this purpose, and each of these projects has proponents extolling the benefits of using Kafka~\cite{datanami, storm-kafka, samza-kafka, flink-kafka, apex-kafka, druid-kafka}.  We therefore use Kafka directly to coordinate communication between our containers.

Apache Kafka implements the publish-subscribe pattern of communication~\cite{birman1987exploiting}.  This pattern operates by relying on an intermediate staging area for data, instead of data producers sending data directly to data consumers.  The intermediate staging area is divided into \textit{topics} to distinguish different kinds of data.  Data \textit{producers} publish data (send it to the intermediate staging area) without any knowledge of the data \textit{consumers} or even a guarantee that any consumers exist.  They simply publish the data as a \textit{topic}.  Similarly, consumers subscribe to a topic, oblivious to all details about the producers.  As producers publish data to a given topic, the data become visible to the consumers subscribed to the same topic; consumers are then free to process the data.

In Kafka, the intermediate staging area where topic data is stored is implemented in \textit{brokers} (these are the brokers mentioned earlier being deployed in their own containers).  A topic is implemented by creating partitions---open file handles---typically spread across multiple brokers.  When a producer publishes data to a topic, it may send that data to any of the partitions, which the corresponding broker receives and writes to the open file.  When a consumer requests data from a partition, the broker reads it from the same file.  In contrast to producers, partitions may have a maximum of one consumer.  Thus an application should divide a topic into at least as many partitions as there are consumers in order to maximize parallelism.

The topic partition also serves as the basic unit of replication.  Kafka assumes and encourages data replication for reliability should a broker go down.  Each partition has a ``leader'' and, in the presence of replication, some number of followers.  After new data is written to a leader partition it is replicated to the followers.  Producers and consumers interact with the broker that holds the leader partition, while the follower partitions are spread among the remaining brokers.  In the event of a broker failure, one of the follower partitions will become the new leader partition.  Unlike partitions, there are no ``leader brokers'' or ``follower brokers''; both leader and follower partitions are spread among all available brokers; thus, no one broker is more important or heavily utilized than any other.

\begin{figure}[t]
    \centering
    \includegraphics[width=0.7\textwidth]{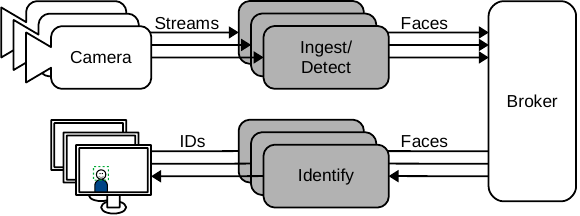}
    \caption{Data center deployment of \application{Face Recognition}.  Algorithms run as standalone processes in lightweight Docker containers deployed on separate nodes throughout the data center.  \textit{Ingestion} and \textit{face detection} both run in the \emph{ingest/detect} container while \textit{feature extraction} and \textit{classification} are combined in \emph{identify.}  Communication between steps within a container happens internally while communication between containers is coordinated through Apache Kafka \textit{brokers}.}
    \label{fig:fr_flow_dc}
\end{figure}

Our data center deployment of \application{Face Recognition} is depicted in \Fig{fig:fr_flow_dc}.  The ingest/detect containers function as producers, sending face thumbnails extracted from each frame to brokers as the ``faces'' topic.  The identification containers are the corresponding consumers, subscribing to the ``faces'' topic.  The placement of brokers between ingest/detect and identification containers was chosen to provide load balancing.  As we will show in \Sec{sec:aitax}, the two containers have different latencies; we thus instantiate more identification than ingest/detect containers.  By placing brokers between them, Kafka ensures that the work is spread among the consumers evenly.
Each of the three container images is deployed a set number of times and distributed throughout the data center.  Because of the extremely low network utilization relative to capacity (\Sec{sec:accelerate:bw}), the placement of containers relative to one another in the data center is unimportant.  
We use a minimum of three broker nodes in all cases to allow for three-way data replication, reflecting common practice in industry-quality deployments.


When we experimented with the three-stage (four containers: ingestion, face detection, identification, and brokers) setup, the communication of video frames between ingestion and face detection was also passed through the brokers.  We simply created an additional topic---frames---within the same set of brokers.


\subsection{Container Resources}
\label{sec:application:container}

\begin{figure}
    \small
    \centering
    \includegraphics[width=0.7\textwidth]{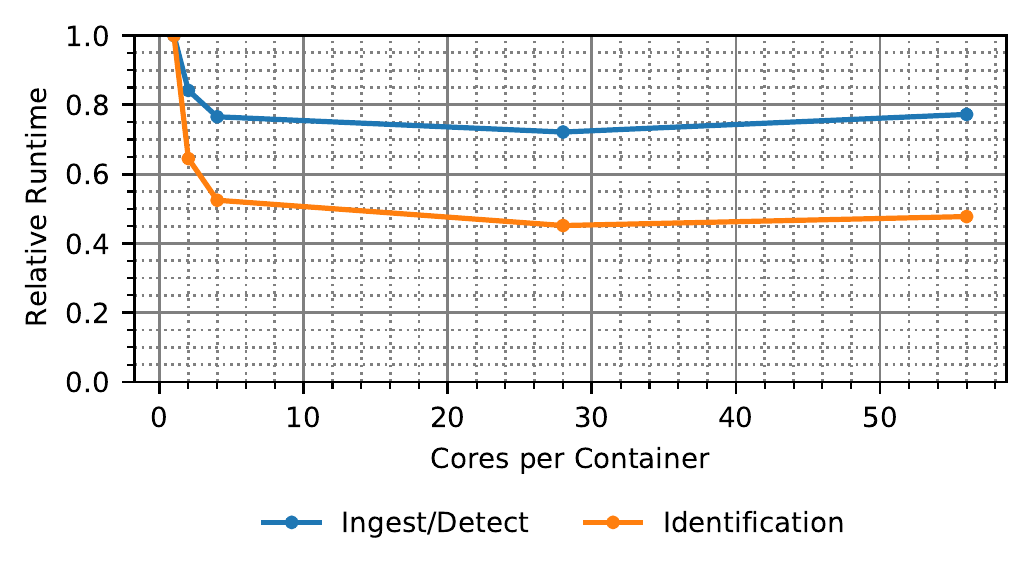}
    \caption{Relative computational latency of \application{Face Recognition} containers with core scaling.  Core scaling shows very limited effectiveness in decreasing the latency, particularly in identification containers.}
    \label{fig:fr_corescaling}
\end{figure}

An important consideration in deploying containers is determining what resources each should be allocated.  \Fig{fig:fr_corescaling} shows how the ingest/detect and identification containers perform with increasing core counts.  With more cores, both containers complete their operations at lower latency, but they do not scale linearly.  Doubling the core count from one to two yields only a 16\% reduction in latency in ingest/detect and a 36\% reduction in identification.  At larger core counts, the computational latency actually increases for both containers.

The ``correct'' allocation of cores to containers depends on the workload requirements.  Often in a data center application the key metric is latency-bounded throughput: maximize the throughput of the application as long as the latency of individual queries remains below some upper bound.  This is a prominent metric used in MLPerf Inference~\cite{janapareddi2019mlperf}.

For \application{Face Recognition}, where we lack a clear latency bound and given the very poor core-scaling behavior, we optimize for throughput by assigning a single core to each container and arbitrarily declare that the resulting latency is acceptable.  This may not be the case in other applications and deployments.

\subsection{Container Ratios}
\label{sec:application:ratios}

As discussed in \Sec{sec:application:stages}, identification can take a variable amount of time, depending on the number of faces that need processing.  Furthermore, we will see in \Sec{sec:aitax} that identification takes significantly longer to perform its calculations than ingest/detect, even when identifying only a single face.  To ensure that identification can keep up with ingest/detect, we allocate many more identification than ingest/detect containers.

The precise ratio of identification to ingest/detect containers is dependent upon the characteristics of the video streams and the latency bound.  If video streams never contain more than one face at a time, fewer identification containers are needed.  However, even if video streams tend to show low face counts on average but have large spikes where there are many faces at once, this can temporarily overwhelm the identification containers and so requires that more such containers are instantiated.  We will see this in our experiments (\Sec{sec:aitax:latency_breakdown}).

\section{AI Tax}
\label{sec:aitax}

We start our exposition of the AI tax by evaluating the end-to-end performance of \application{Face Recognition}.
We aim to understand what fraction of the cycles in an AI application go to AI processing versus the non-AI components.  To this end, we examine the lifetime of a frame as it flows through the AI-centric \application{Face Recognition} application.  In \Sec{sec:aitax:events}, we explain how we measure frame progress.  \Sec{sec:aitax:latency_breakdown} breaks down the end-to-end progress of a frame in each stage of the pipeline and shows that AI computation is not so central as one would expect in an AI application.  In \Sec{sec:aitax:process} we break down the application behavior in each container and reveal how much supporting compute is needed to enable AI processing.  We show that it is vitally important to view AI application performance holistically, as it involves much more than just AI processing and the supporting code and infrastructure tax have a profound impact on latency.  

\subsection{Instrumenting the End-to-End Execution}
\label{sec:aitax:events}

\input{tex/code/events}

To really understand an AI application deployed in even an edge data center, we must raise the level of abstraction from how applications are traditionally evaluated.
While we do not claim to have the right level of abstraction for all end-to-end workloads, for \application{Face Recognition}, we believe that we have identified a good level of abstraction for tracking and measuring application progress without perturbing the application's original behavior.

Application progress is a sequence of unit steps that are necessary for a frame to progress through the application.  We term the units of application progress ``events''; these are high-level steps in the application and correspond to the stages described in \Sec{sec:application}: video ingestion, face detection, broker waiting time, and identification.  \Lst{lst:events} shows simplified Python code demonstrating the operation of the face detection process with event-logging code inserted.  The event-logging code in the listing is only slightly simplified from our actual code.  Event-based logging lets us track end-to-end application progress.  This higher level of abstraction is critical in enabling engineers to architect at a cluster level, where the complete application executes, instead of just at a node level.

We log all the events during execution of the application using Elasticsearch~\cite{elasticsearch} and Logstash~\cite{logstash} running on a separate server.  We measure the execution time of each step as well as the sizes of data that are transferred between stages.  This is done using timestamps around the major regions of interest, e.g.\ the time to do the face identification \textit{excluding} the supporting code (e.g.\ iteration management).  In essence, events capture the major steps that a frame goes through from beginning to end.  We use built-in language functions to measure the size of the data that are transferred through the brokers.  Due to the infeasibility of instrumenting a complex program such as Kafka, we approximate the broker waiting time event by calculating the time delta between the end of face detection and the beginning of identification.

\begin{figure}
    \centering
    \small
    \begin{subfigure}{0.28\textwidth}
        \includegraphics[width=1.0\textwidth]{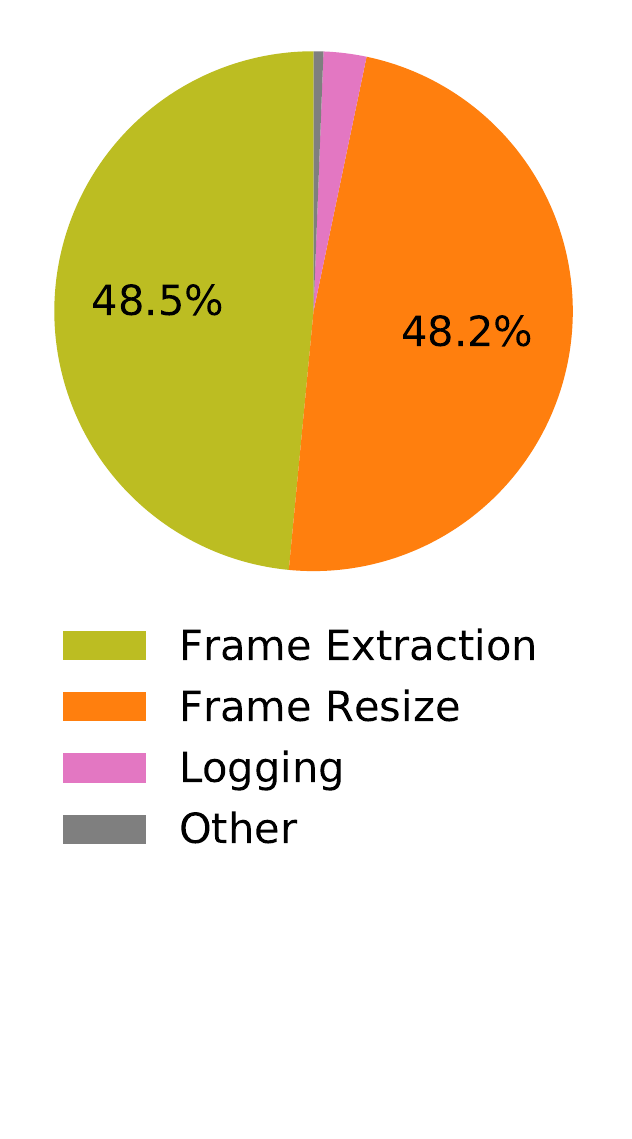}
        \caption{Ingestion.}
        \label{fig:process_breakdowns:ingestion}
    \end{subfigure}
    \hspace{2ex}
    \begin{subfigure}{0.28\textwidth}
        \includegraphics[width=1.0\textwidth]{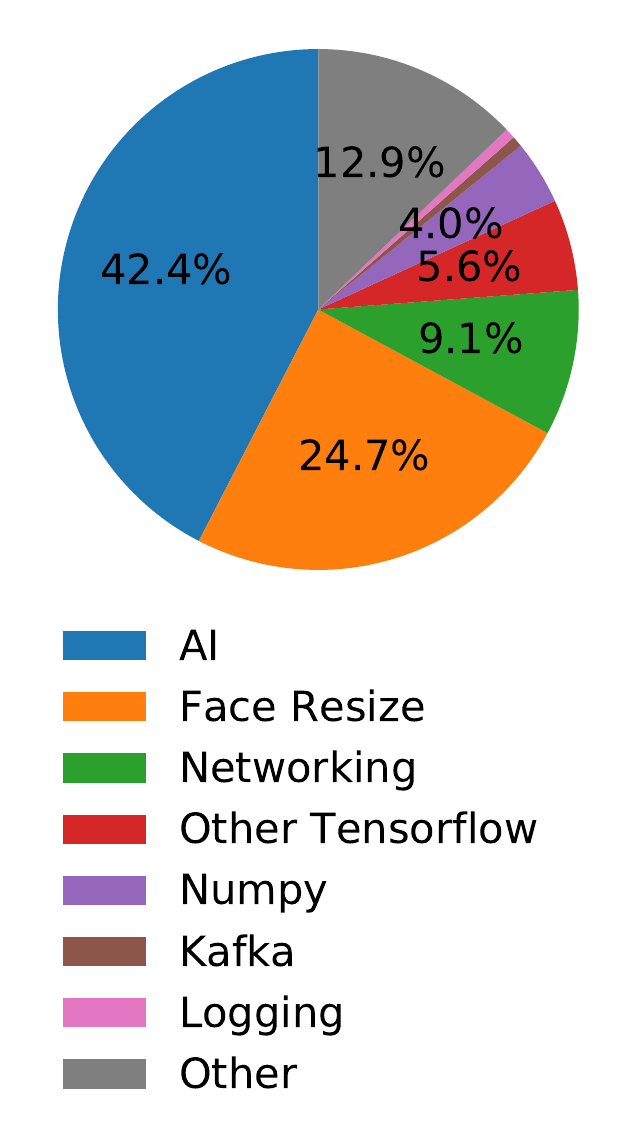}
        \caption{Detection.}
        \label{fig:process_breakdowns:detection}
    \end{subfigure}
    \hspace{2ex}
    \begin{subfigure}{0.28\textwidth}
        \includegraphics[width=1.0\textwidth]{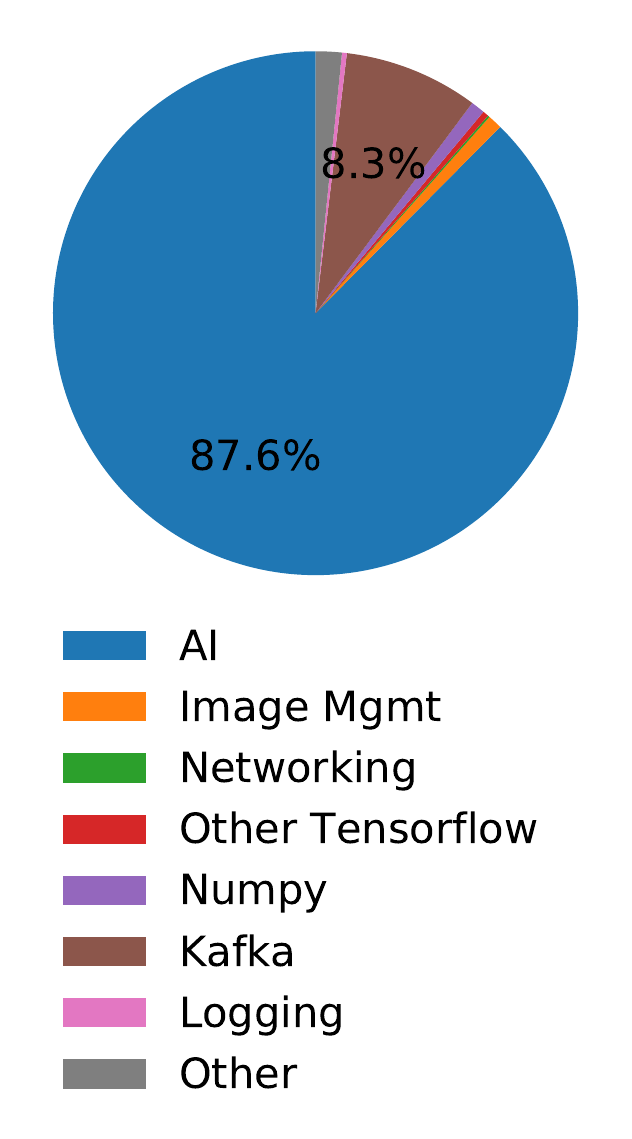}
        \caption{Identification.}
        \label{fig:process_breakdowns:identification}
    \end{subfigure}
    \caption{Process CPU time breakdowns.  Though ingestion uses no AI algorithms, it consists of straightforward image processing algorithms, which should be easily accelerated~\cite{lindoso2009hardware}.  Face identification is overwhelmingly AI-centric.  In contrast, face detection relies heavily on supporting code to enable its AI processing.}
    \label{fig:process_breakdowns}
\end{figure}

Our instrumentation method has negligible overhead and resource requirements, since we are only logging events.  It has minimal impact on the application (see \Fig{fig:process_breakdowns}).

\begin{figure}
    \centering
    \includegraphics[width=0.65\textwidth]{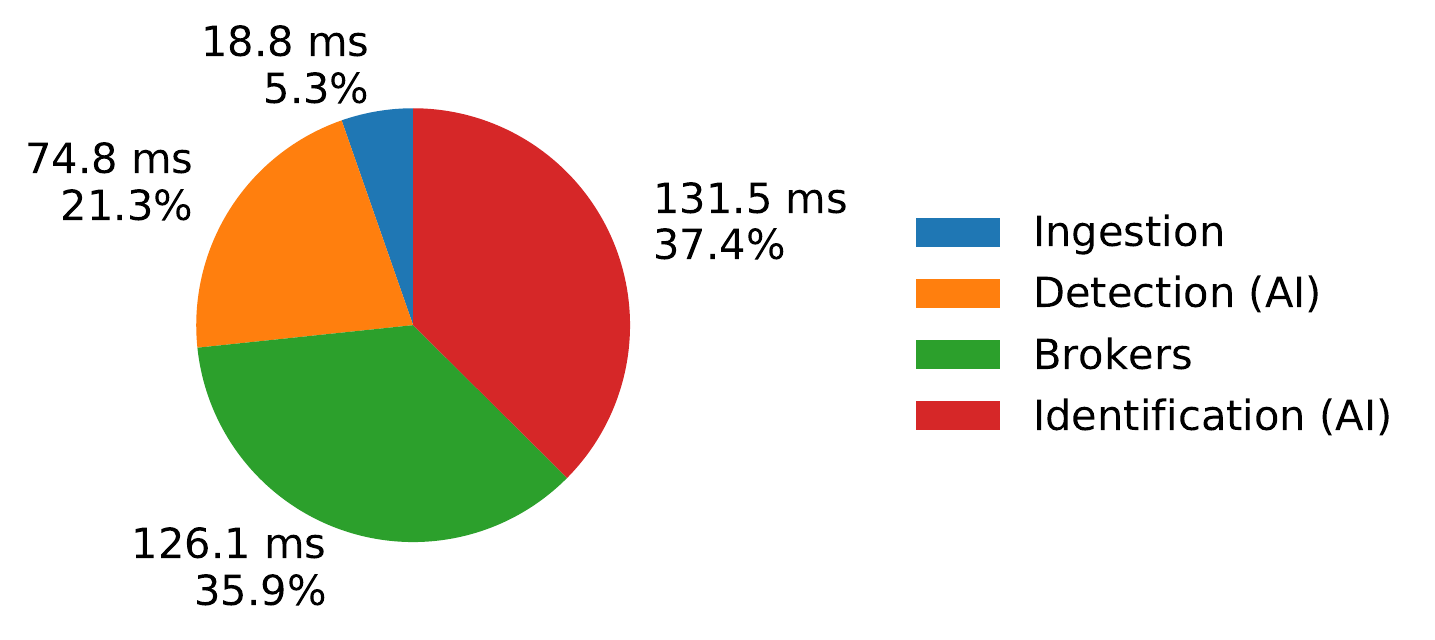}
    \caption{Breakdown of end-to-end frame latency.  With inference steps in detection and identification stages, less than 60\% of the latency arises from AI stages.  Over a third of a frame's lifetime is spent in brokers.  
    }
    \label{fig:fr_latency_pie}
\end{figure}

\subsection{AI Applications Are More Than Just AI}
\label{sec:aitax:latency_breakdown}

One of the end user's primary concerns is latency.
In \application{Face Recognition}, this is the total time of a frame progressing serially from ingestion through identification; the latency of any stage that is not performing AI contributes to the AI tax.

We conduct experiments with 840 ingest/detect processes (producers) executing on 15 nodes (56 processes per node), 1680 identification processes (consumers) executing on 30 nodes (56 processes per node), and 3 brokers (each given its own node).  We require the brokers to maintain 3$\times$ data replication, which is standard practice for disaster recovery.

We measure an average face size of 37.3~kB and an end-to-end latency of 351~ms.  While this latency may seem large, there are two points to remember.  First, the throughput per stream is around 10 frames per second (FPS)---and a single stream could be divided among three ingest/detect instances for 30 FPS operation---regardless of the latency, since the application is pipelined; the output video still displays smoothly, just with a small delay.  Second, there are multiple inferences per frame, performed sequentially, with the inference stages located on different nodes to improve performance~\cite{gupta2019architectural}; the communication between the stages imposes additional latency.

\Fig{fig:fr_latency_pie} summarizes the average latency for each stage.  Ingestion operates quickly, taking only 18.8~ms, while the AI stages, face detection and identification, take 74.8 and 131.5~ms, respectively.  Remarkably, over a third of the end-to-end latency is spent waiting between stages, at 126.1~ms.  As in any real-time application, tail latency is an important factor to consider.  We measure a \nth{99} percentile tail latency of 2.21~s, with the standalone \nth{99} percentile tail latencies of ingestion, detection, waiting time, and identification at 27~ms, 1.84~s, 116~ms, and 380~ms, respectively.  We remind the reader that these latencies can be improved somewhat by allocating additional cores to each stage of processing; in our implementation, however, we have chosen to optimize for throughput over latency.



\begin{figure}[t]
    \small
    \centering
    \includegraphics[width=0.8\textwidth]{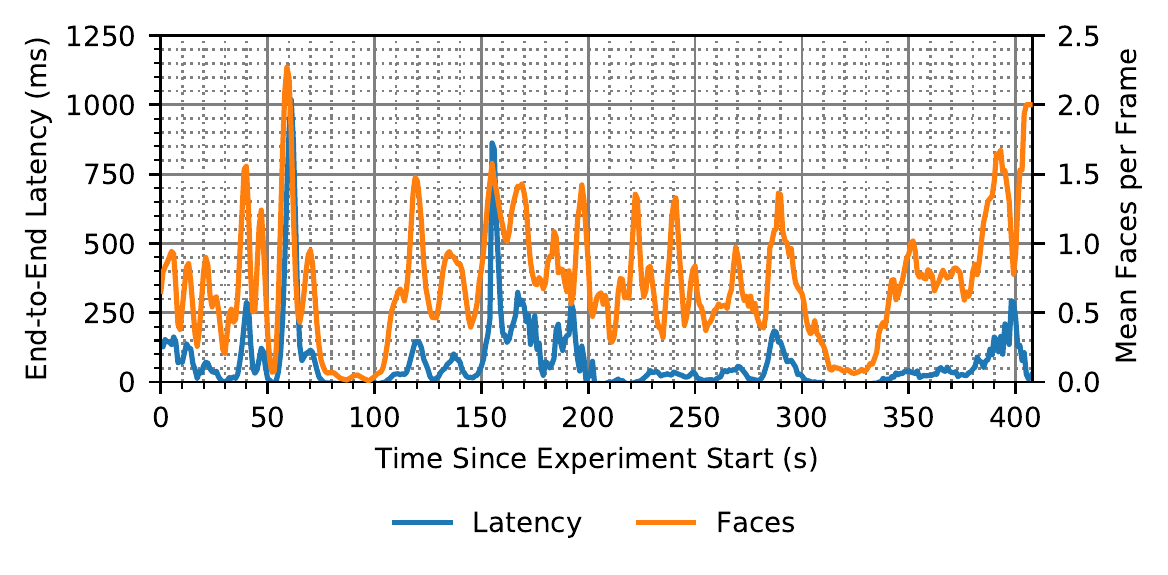}
    \caption{Latency tracks the total number of faces in the system.  As the experiment runs, average end-to-end latency is clearly correlated to the number of average faces per frame; with higher average faces per frame, the total number of faces in the system is pushed upward, causing more delay through the communication system.}
    \label{fig:fr_latency_timeline}
\end{figure}

From the tail latency breakdowns, we see that the end-to-end tail latency derives almost entirely from the waiting time in the brokers.  This waiting time in turn results, at least in part, from congestion in the application.  As shown in \Fig{fig:fr_latency_timeline}, when ingest/detect processes collectively produce a surplus of faces, identification has a hard time keeping up, leaving the faces in the brokers for a longer time.  When there are almost no faces detected, identification containers are almost idle and so are able to fetch the few faces quickly.

In summary, deep learning inference performance is more than just the performance of an individual node in the system. Even with a well-balanced system, there is a substantial AI tax latency imposed in managing the transfer of data between the nodes (i.e.\ the detection and identification stages).  Without looking at the end-to-end latency, one would not realize that a large portion of time is spent waiting at brokers.  This observation is \emph{not} unique to our application; any application built on Apache Kafka (or a similarly brokered communication mechanism) will face this reality.

\subsection{Overhead of Pre- and Post-Processing AI}
\label{sec:aitax:process}

Most AI research papers focus only on the core AI component, neglecting the other associated parts that are essential to end-to-end AI processing. However, there are pre- and post-processing steps that are unavoidable. Both play a critical role in the overall latency and both contribute to the AI tax. Pre-processing involves preparing the data for the AI kernel execution, while post-processing is loosely defined as any processing that is performed to convert the generated AI result(s) into something meaningful to the user or next stage.

To quantify the AI tax for pre- and post-processing we refine our view of a frame's processing by looking at the time breakdown of each process using code profiling tools.  \Fig{fig:process_breakdowns} shows, for each of the main processes (ingestion, detection, and identification), where time is spent.

Ingestion is exclusively a pre-processing stage. It shows a nearly even split between frame extraction and frame resizing  (\Fig{fig:process_breakdowns:ingestion}).  Extraction refers to parsing the incoming video stream into individual frames.  Resizing converts frames from \texttt{1920x1080} to \texttt{960x540} for the detection stage.  The remaining time is split between the overhead of event logging and other supporting code, including transferring frames to the co-located detection process.

During face detection (\Fig{fig:process_breakdowns:detection}), despite being an AI-centric stage, only 42\% of the time is spent executing the AI algorithm in TensorFlow.  Cropping and resizing faces (to \texttt{160x160}) for identification takes 25\% of the time; supporting TensorFlow and NumPy code (pre- and post-processing for each frame) take 6\% and 4\%, respectively; and ``other'' code takes a whopping 13\% of the time.  Code in the ``other'' category includes inter-process communication (from the ingest stage), additional matrix manipulation, loop management, bounding box calculation, image encoding, etc.

The AI-centric identification stage has a markedly different breakdown.  It spends 88\% of its time directly executing AI algorithms; Kafka code, though, takes 8\% of the time.  The remaining components contribute little to the total time.

Beyond the pre-processing of the ingestion stage, end-to-end \application{Face Recognition} requires substantial pre- and post-processing within the AI-centric stages.  In face detection, non-AI computation constitutes 57.6\% of the compute cycles.  In identification, that figure drops to 12.4\%, which is still far from trivial.  In a complex and diverse AI-centric application such as \application{Face Recognition}, AI computation constitutes 55.2\% of end-to-end cycles, with the remainder going to supporting code: 17.8\% to resizing, 9.0\% to networking, 5.2\% to tensor preparation, 3.6\% to Kafka processing (outside of the brokers), and the rest to other supporting tasks.

In summary, despite the massive excitement surrounding AI algorithms, AI workloads are more than just tensors and neural networks: without the supporting code, AI is impotent.  We emphasize that the supporting code, far from being a minor player in a complete application, constitutes over 40\% of the compute cycles, not counting the compute time in the brokers. The pre- and post-processing code is executed on the general-purpose CPU, so it motivates the need to understand the role of the CPU as AI acceleration increases.

%
%
%

\section{Accelerating AI and Its Implication on AI Tax}
\label{sec:accelerate}

There are numerous efforts underway to accelerate AI~\cite{aichip, habana, chen2014diannao, intelncs2, mahajan2016tabla, xilinx2018alveo, nvdla, googlecoral}.  But given the significance of the AI tax in end-to-end AI performance and in pre- and post-processing, it is important to understand how the tax evolves as AI is accelerated and its impact on the overall end-to-end application performance; there are performance limitations that arise beyond a certain point of AI acceleration.  To build a balanced system, it is important to understand these limits.  

We therefore study how varying degrees of AI speedup affect the end-to-end performance of our video analytics workload.
In general, we could foresee accelerated AI coming about in various ways: CPU manufacturers may decide to integrate ML-centric hardware directly into the CPU execution pipeline; or dedicated off-chip accelerators may be utilized, including highly-parallel pipelines (such as GPUs) and dedicated inference engines (such as Intel's Neural Compute Stick~\cite{intelncs2}, Habana's Goya inference processor~\cite{habanagoya}, or Google's Coral Edge TPU~\cite{googlecoral}).  Comparisons of the efficacy of each of these solutions is the subject of other work.  In this work, we look at the impact of theoretical speedups, regardless of how the speedups are achieved.

In this section, we explore the impact of accelerating AI applications up to 32$\times$, based on the performance of existing accelerators.  Habana reports that its Goya processor achieves 13.5$\times$ speedup over a two-socket Intel Xeon Platinum 8180~\cite{wheeler2018data}.\footnote{Other than its faster clock, the 8180 CPU is identical to the 8176 CPUs we use.}  We explore speedups approximately twice as great as this (up to 32$\times$) in order to account for future advances.

\Sec{sec:accelerate:process} analytically estimates accelerated AI performance to show its asymptotic limits.  \Sec{sec:accelerate:now} introduces our technique for emulating accelerated workloads on current hardware.  In \Sec{sec:accelerate:impact} we evaluate the performance of accelerated processing and discover a quickly approaching bottleneck.  \Sec{sec:accelerate:bw} shows that the bottleneck results from overwhelming the system's capacity to write to storage.  We finish by showing in \Sec{sec:accelerate:waiting} how frames' waiting time in brokers grows as a fraction of end-to-end latency.

\subsection{Analytical Speedups for AI Acceleration}
\label{sec:accelerate:process}

\begin{figure}
    \small
    \centering
    \includegraphics[width=0.7\columnwidth]{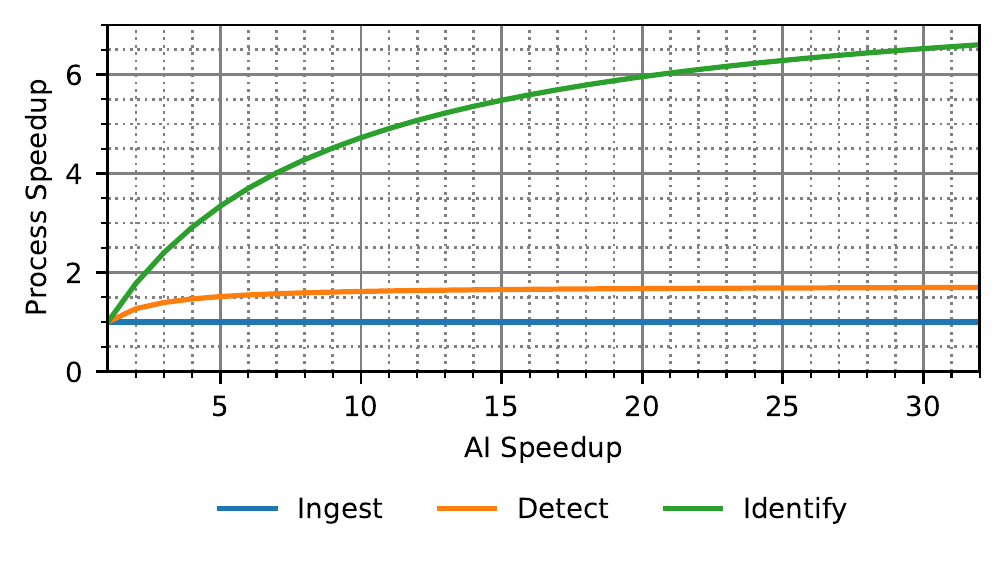}
    \caption{Projected speedups of individual processes.  Amdahl's law predicts that the realized speedups of accelerated AI diminish quickly, approaching an asymptote far short of the speedups of AI accelerators.}
    \label{fig:fr_process_profiles_model}
\end{figure}

The AI tax means that a significant portion of an AI application's compute cycles are spent on tasks other than AI and ML, and Amdahl's law dictates that the overall speedup of a system is limited by the portion of execution that is not accelerated.
Application of Amdahl's law (\Fig{fig:fr_process_profiles_model}) shows that each of the three primary processes---ingestion, detection, and identification---is limited in how much real-world speedup it can enjoy if AI is accelerated in isolation.
Ingestion, which performs no AI compute, naturally derives no benefit from acceleration.  Detection, which is 42\% AI, rapidly approaches its asymptotic speedup of just 1.74$\times$, achieving 1.59$\times$ overall speedup at 8$\times$ acceleration and 1.66$\times$ overall speedup at 16$\times$ acceleration.  Identification, at 88\% AI, has an asymptotic speedup limit of just 8$\times$.  At 16$\times$ AI acceleration it achieves 5.6$\times$ overall speedup, and even at 32$\times$ AI acceleration it shows just 6.6$\times$ overall speedup.

The exciting speedups promised by up-and-coming inference accelerators will be severely moderated by the reality of the supporting, non-AI code---the AI tax.
With the fervor surrounding acceleration of AI and ML, these results from Amdahl's law serve as an important reminder that AI applications are more than ML computation.  Supporting and enabling code is a critical component of an end-to-end application and this should serve as a call to action to address the limitations imposed by that code.

\subsection{Emulating AI Acceleration on Hardware}
\label{sec:accelerate:now}

It is instructive to see how the AI tax evolves as compute is universally accelerated (i.e.\ overcoming the asymptotic limits of \Sec{sec:accelerate:process}) on a real system.  To do so, we emulate the behavior of accelerated processing.  Only the most basic loop controls and Kafka code are left in their original state.

Our emulated acceleration technique relies on the observation that, from the perspective of application progress, the perspective of network traffic, and the perspective of the brokers, it is impossible to distinguish between (1)~running the real application as has been described and characterized and (2)~implementing artificial delays reflective of the actual compute times (\Sec{sec:aitax}) and sending meaningless data over the network of the same size as in the real application.  In accelerated \application{Face Recognition}, rather than accelerating and executing the real algorithms, we replace the compute with calls to \texttt{sleep}, where the sleep duration is reflective of measured execution times (\Sec{sec:aitax:events}).  Accordingly, rather than sending face thumbnails to brokers, we send meaningless data whose size matches the measured sizes.
We can accelerate processing (both ingest/detect and identification) by an arbitrary factor by dividing the sleep times by the speedup factor.  In this way, we maintain the behavior of the brokers, network, storage, and supporting code while exploring how acceleration changes the AI tax.

\begin{figure}
    \centering
    \small
    \includegraphics[width=0.7\textwidth]{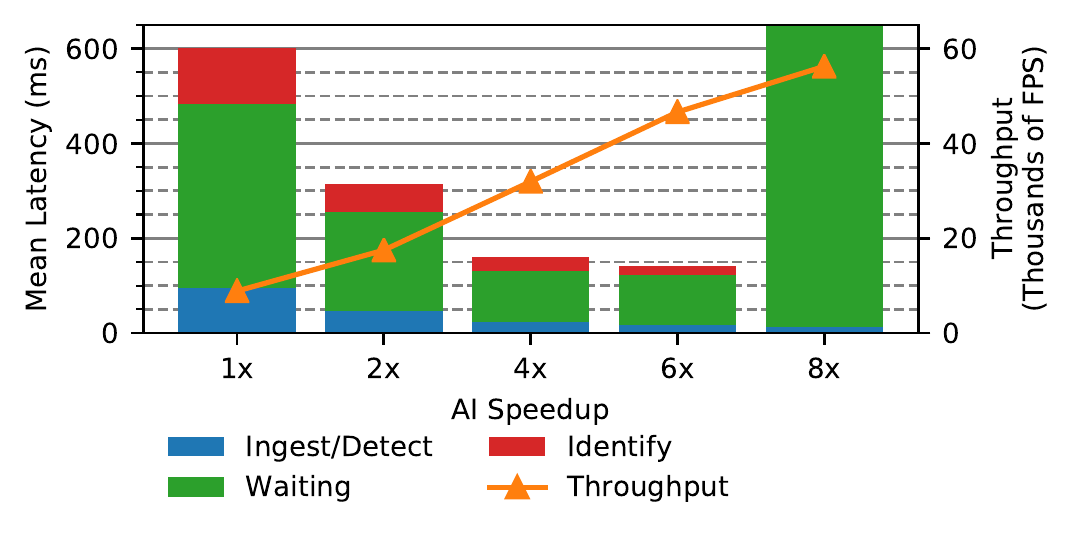}
    \caption{Average frame latency and throughput under increasing AI acceleration.  Beyond 8x speedup, the increased throughput leads to an unbalanced system.  Queueing theory dictates that if elements enter the system faster than they leave, the latency increases to infinity.}
    \label{fig:fr_speedup_frame}
\end{figure}

\begin{figure}
    \centering
    \small
    \begin{subfigure}{1.0\textwidth}
        \centering
        \includegraphics[width=0.7\textwidth]{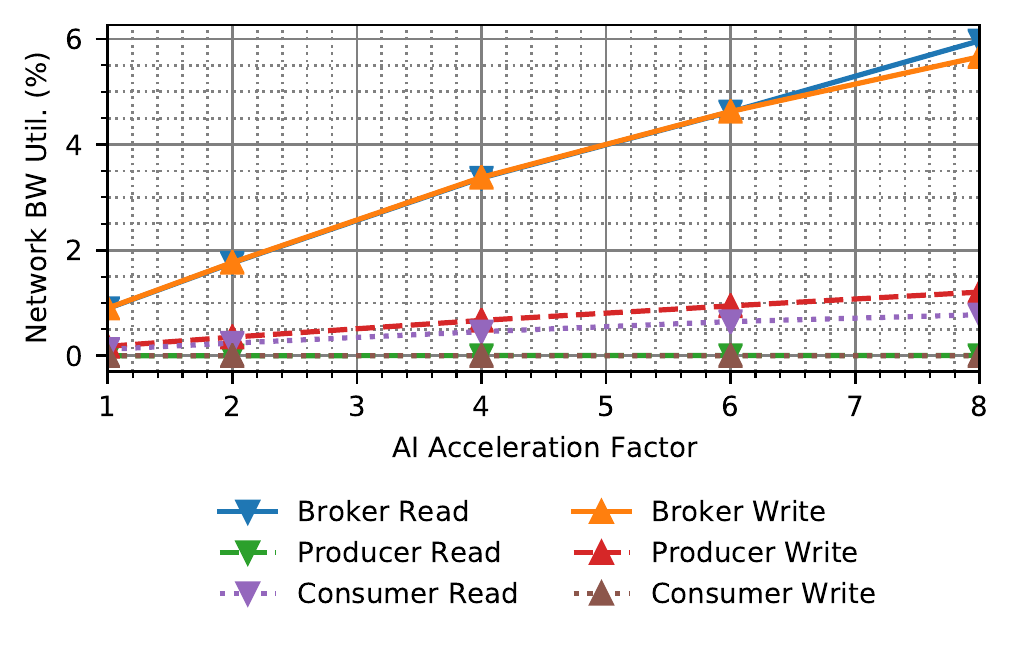}
        \caption{Network bandwidth utilization.  Network activity from and to producers and consumers is concentrated at the brokers; nevertheless, broker network utilization falls far short of our 100~Gbps capacity.}
        \label{fig:fr_speed_bw_util:network}
    \end{subfigure}

    \vspace{1em}
    \begin{subfigure}{1.0\textwidth}
        \centering
        \includegraphics[width=0.7\textwidth]{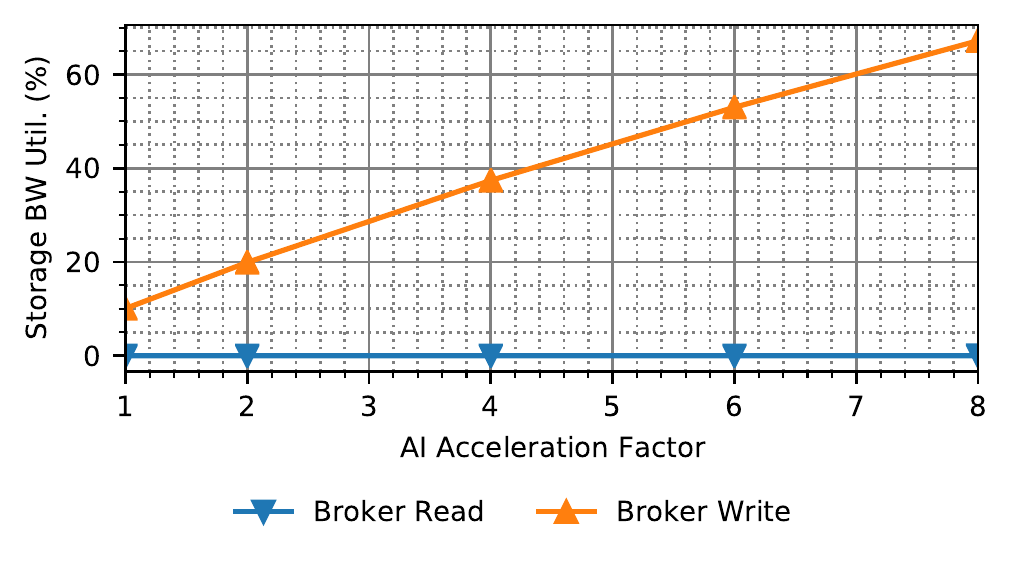}
        \caption{Storage bandwidth utilization.  Storage activity for producers and consumers is not shown because it is not preserved in accelerator emulation.  Network activity is translated to storage activity in the brokers.}
        \label{fig:fr_speed_bw_util:storage}
    \end{subfigure}
    \caption{Network and storage bandwidth utilization under acceleration.  Whereas network utilization never exceeds 6\% of network capacity, storage bandwidth utilization exceeds 67\% of capacity at 8x acceleration.  Storage bandwidth becomes a bottleneck much sooner than network bandwidth.}
    \label{fig:fr_speed_bw_util}
\end{figure}

We emphasize that this AI acceleration emulation provides realistic performance estimation under acceleration because (1)~the most basic general purpose processing (the support code to iterate through available frames, code to coordinate communication with Kafka brokers, the brokers themselves, etc.) remains in place and is executed as usual without the benefits of acceleration; (2)~from the perspective of the data center, compute time spent executing real algorithms and waiting in \texttt{sleep} are identical; and (3)~the brokers are completely ignorant of and unconcerned with the execution details of both producers and consumers.  Thus, our setup to accelerate AI through emulation provides a realistic and believable look at the impact of faster AI on the data center and on the workload as a whole.

\subsection{Accelerated AI Impact on Total Speedup}
\label{sec:accelerate:impact}

We explore how the end-to-end frame latency will evolve as AI benefits from increasingly powerful acceleration.  For this analysis, we assume that the AI algorithms will experience no latency overhead from acceleration---that is, we assume that the latency to communicate with dedicated off-chip accelerators is factored into the emulated speeds, or, equivalently, that future CPU architectures will directly integrate accelerator hardware into their execution pipelines.

For these experiments, we maintain the same application organization as depicted in \Fig{fig:fr_flow_dc}.  As we accelerate producers and consumers equally, the imbalance between the two will persist, so it is important to utilize Kafka's load balancing features at the interface between the two.

For the sake of simplicity and repeatability and without loss of generality, we configure these emulation experiments so that each frame produces exactly one face.  This has two impacts on performance: (1)~on average, this is more faces per frame than produced by the video file used previously, which yields 0.64 faces per frame on average; and (2)~because the rate of face production is constant, we do not have to provision our cluster to handle sudden spikes in traffic, allowing us to deploy fewer identification instances than for the video file.  None of the conclusions we draw from these experiments is invalidated by these these two observations.

\Fig{fig:fr_speedup_frame} shows the effects of accelerating the AI components of \application{Face Recognition}.  Note that because we assume one face per frame, which is significantly higher than the 0.64 faces per frame average produced by our default video file, the average end-to-end latency is somewhat higher at 1$\times$ speed than in \Sec{sec:aitax:latency_breakdown}.  At higher speedups, we see a two-fold benefit: first, the latency is very clearly reduced; second, the throughput is commensurately increased.

At 8$\times$ speedup, we see a new manifestation of the AI tax, with latency tending toward infinity---the longer the experiment runs, the larger the latency grows.  This is an example of an unstable system in queueing theory: faces are entering the system more quickly than they are leaving.  This is a major limitation that can severely hamper the prospects of AI acceleration and demands further investigation.

\subsection{Network and Storage Bandwidth Limits}
\label{sec:accelerate:bw}

With state-of-the-art industry accelerators claiming improvements of up to 15$\times$ in inference speedup over CPUs~\cite{habana}, it is critical to understand why the system becomes unbalanced at 8$\times$ acceleration.  Without this insight, it will be difficult to build systems to accommodate higher acceleration factors.

Intuitively, we suspect the imbalance results from the increased throughput of the system overwhelming one of two resources with limited bandwidth: either network or storage bandwidth.  We measure the utilization of both bandwidths to understand the problem at increased acceleration factors (i.e.\ greater than 8$\times$).  In \Fig{fig:fr_speed_bw_util:network}, the network bandwidth utilization of all container types rises with increasing acceleration factor.  Unsurprisingly, producer (ingest/detect containers) network read bandwidth is next to zero, as is consumer (identification containers) write bandwidth.  Conversely, producer write and consumer read bandwidths are comparable.  But the real network bandwidth hot spot is the brokers---as the point of communication between producers and consumers, they must process all network traffic generated by the producers or read by the consumers.  However, even the combined network traffic flowing through the brokers constitutes a small portion of the available bandwidth: at 8$\times$ accelerated AI, the read bandwidth is only 6~Gbps, a mere 6\% of the available 100~Gbps.

\Fig{fig:fr_speed_bw_util:storage} shows the storage bandwidth requirements of the brokers.  We omit the data for the producer and consumer containers, as their storage behaviors are not preserved by our emulation technique and are nevertheless expected to be near zero, as they work largely out of memory.  The brokers, however, have rather high bandwidth requirements.  Even at native (1$\times$) speed, the write bandwidth is 10\% of capacity (1.1~GB/s).  At 8$\times$ acceleration, that rises to over 67\%.

With the overhead of the operating system, managing the file system, and coordinating all the small requests to be written to storage, by 8$\times$ acceleration, 67\% utilization has effectively saturated the available bandwidth.  Returning to queueing theory, the inability of storage to write data to storage (and make it available to the consumers) as fast as it is supplied leads to the imbalance and growing latency.

We note that data reads, however, use essentially none of the available bandwidth.  This is easily understood: brokers are tasked with ensuring data reliability, so they must write producer data to storage, but the operating system can also cache the data in memory, allowing reads directly from memory and bypassing the storage read path.

Doubtless, fine-tuning the brokers' parameters could allow them to better utilize the storage bandwidth.  An in-depth exploration of the Apache Kafka parameter space is not, however, the purpose of this paper.  Regardless of the ability of the brokers to utilize available bandwidth, they will hit a hard limit at the specifications of the hardware devices.

In a setup with a more conservative network bandwidth (e.g.\ 10~Gbps), both the storage and the network would quickly become bottlenecks when accelerating compute.

Thus the increased throughput of a moderately accelerated end-to-end system creates a new AI tax that quickly overwhelms the communication substrate, counteracting the gains achieved through hardware acceleration of AI.

\subsection{Increase in Waiting Time}
\label{sec:accelerate:waiting}

Furthermore, whereas the waiting time at 1$\times$ speed constitutes 64.6\% of the total latency of a frame (\Fig{fig:fr_speedup_frame}), it grows to 66.4\% at 2$\times$, 68.0\% at 4.0$\times$, and 79.1\% at 6$\times$.  This trend can be partially understood by Kafka's automatic batching between brokers and consumers and producers.  A message from a producer can be held in the producer for a small amount of time until a larger group of messages has been accumulated to be sent as a batch.  Similarly, when a consumer requests available messages from a broker, the broker can withhold messages until there exists some minimum amount of data.  Thus, the broker time grows with the decrease in compute time to improve batching.  Both batching behaviors are limited by timeouts to ensure that neither producer nor consumer waits excessively long.  We have tuned these parameters to find settings that ensure good behavior across a variety of experiments.  Nevertheless, the time spent waiting between producers and consumers approaches some lower limit beyond which no amount of tuning can help; in an application that has many more stages than \application{Face Recognition}, this minimum waiting time could accumulate across stages and prove prohibitively long.

\section{Generalizability of Findings}
\label{sec:discussion}

We recognize that our research is a case study of a single application.  While case studies are often undervalued---despite shedding light on an application that is valuable in its own right and pioneering an evaluation approach~\cite{pingali2019case}---we nevertheless acknowledge that evaluation of additional applications is beneficial.  We therefore discuss some findings on a second application, \application{Object Detection}, that was deployed similarly using Kafka in our edge data center.  While we do not study \application{Object Detection} in the same detail as \application{Face Recognition}, we discuss here some results showing that this additional application faces AI tax bottlenecks as well.  In \Sec{sec:discussion:object}, we describe the purpose and design of the application.  We look at the AI tax in \application{Object Detection} when running natively (\Sec{sec:discussion:aitax}) and when accelerated (\Sec{sec:discussion:acceleration}).  


\subsection{Object Detection}
\label{sec:discussion:object}

Like \application{Face Recognition}, \application{Object Detection} analyzes video streams in real-time.  Instead of recognizing faces, though, \application{Object Detection} uses an R-CNN~\cite{ren2015faster} to identify multiple objects in each frame.  Also like \application{Face Recognition}, \application{Object Detection} is split into two stages with Kafka brokers serving to transfer data between them.  The two stages are termed ingestion and detection.  Ingestion ingests a video stream, parsing it into separate frames, and passes those frames through Kafka to detection.  AI compute is exclusively performed in this later stage in the R-CNN.

Unlike \application{Face Recognition}, wherein the presence or absence of faces in a frame dictates whether and how much data is sent through Kafka, in this application each frame is always sent.  This decreases the variability in system load, as each detection instance always has to process precisely one frame at a time.

\begin{figure}
    \small
    \centering
    \includegraphics[width=0.7\textwidth]{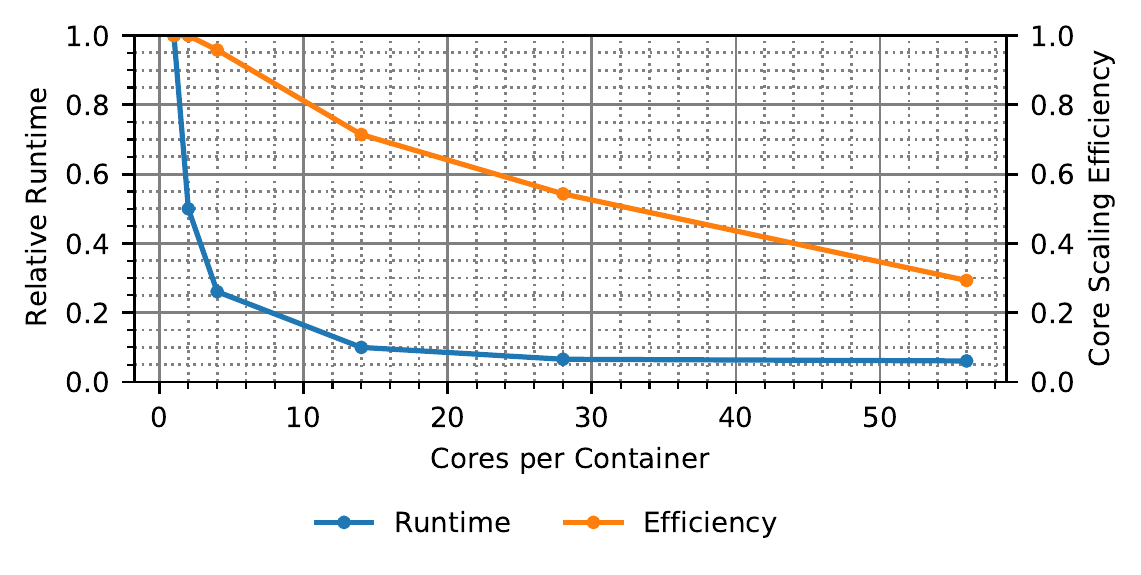}
    \caption{Relative computational latency of \application{Object Detection}'s detection containers with core scaling.  Core scaling shows very good efficiency in reducing runtime, particularly compared to \application{Face Recognition}.}
    \label{fig:od_corescaling}
\end{figure}

As shown in \Fig{fig:od_corescaling}, the detection stage of \application{Object Detection} shows near linear speedups with increasing core count.  Through testing, we determined to allocate 14 cores per container; this allows us to instantiate 4 detection containers per server.  Despite this, the ingestion stage operates orders of magnitude faster than the detection stage (see \Sec{sec:discussion:aitax}); we limit the ingestion rate to 30 frames per second.  To balance that ingestion rate, we instantiate 96 detection containers for each ingestion container.

\subsection{AI Tax}
\label{sec:discussion:aitax}

\begin{figure}
    \centering
    \includegraphics[width=0.65\textwidth]{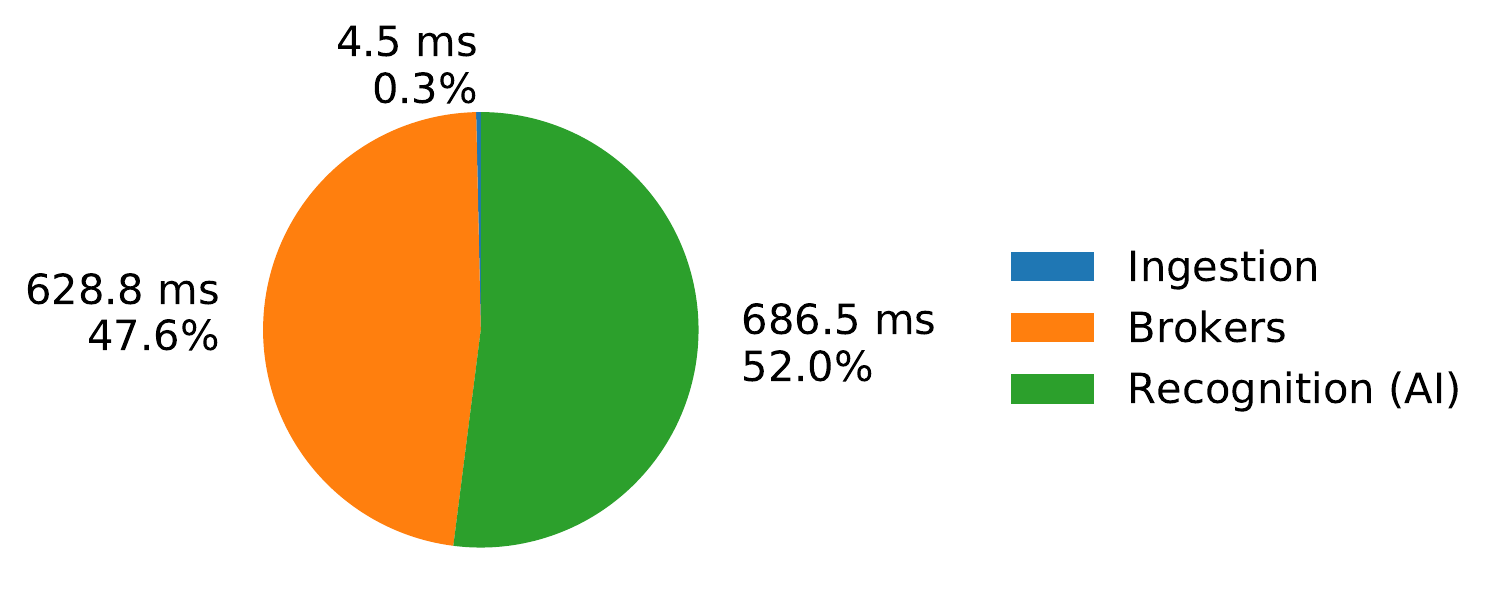}
    \caption{Breakdown of \application{Object Recognition} average end-to-end frame latency.  Unlike \application{Face Recognition}, \application{Object Detection} performs no AI computation in its initial stage (Ingestion); hence, it is orders of magnitude faster than the Recognition stage, where all the AI computation is performed.}
    \label{fig:or_latency_pie}
\end{figure}

\Fig{fig:or_latency_pie} shows the end-to-end frame latency breakdown.  The first stage, ingestion, performs no AI.  As such, it completes very quickly, in only 4.5~ms.  As stated, we limit the ingestion rate to 30~FPS, so for practical purposes this time is really 33.3~ms.  In contrast, the final stage, detection, does all of the AI processing and weighs in at an impressive 687~ms.  Waiting time in the brokers is nearly as long, averaging 629~ms.

\subsection{Acceleration}
\label{sec:discussion:acceleration}

Using our acceleration emulation methodology, we explore the implications of accelerating \application{Object Detection}.  Given the large number of cores needed for each detection instance, and given the limited size of our cluster, we assign only a single core to each detection instance for these experiments, allowing us to deploy significantly more ingestion instances than we would be able to otherwise.  This is feasible because the emulation methodology does not care about number of cores and because in an accelerated data center setup, it may be very possible to support the higher instance count.

We maintain the same ratio of producers to consumers as in our real setup, but by increasing the density of consumers we are able to scale up the experiments significantly.  We use a single producer node but instantiate 21 producers on it and we use 36 consumers nodes, each with 56 consumers.  We continue to use three brokers.

Since we already decided to limit the frame rate to 30~FPS, we continue this practice.  With increasing acceleration, we increase the number of frames we send---effectively, the acceleration factor dictates the number of simultaneous video feeds each producer can process.  Thus, at 2$\times$ acceleration, each producer sends frames at 30~FPS but it sends two frames at a time instead of one; similarly, at 8$\times$ acceleration, a producer sends eight frames at 30~FPS, and so on.

\begin{figure}
    \centering
    \small
    \includegraphics[width=0.7\textwidth]{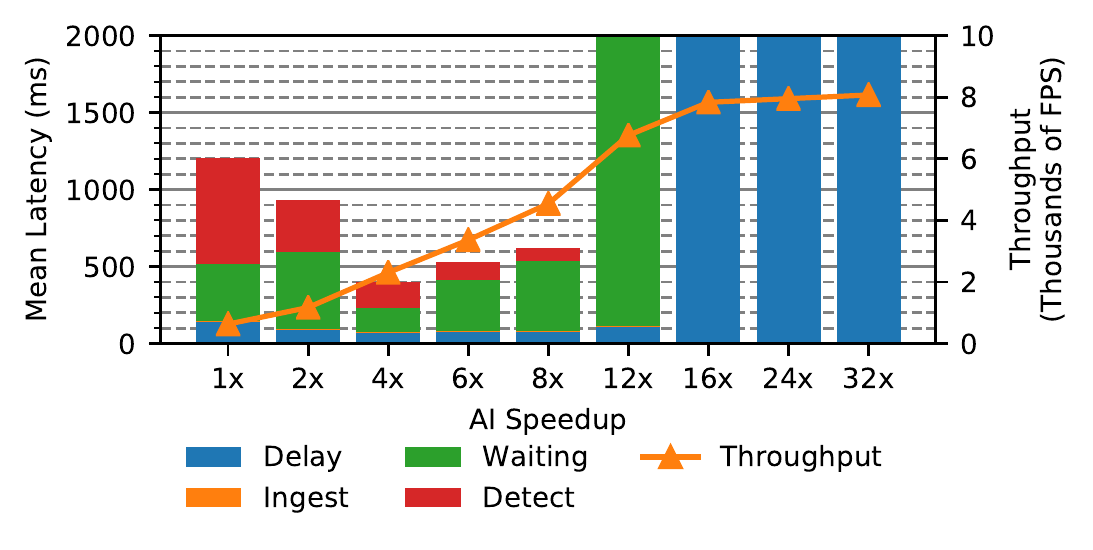}
    \caption{\application{Object Detection} average frame latency and throughput under increasing AI acceleration.}
    \label{fig:or_speedup_frame}
\end{figure}

\Fig{fig:or_speedup_frame} shows the results of acceleration.  Despite the significantly lower count of producers compared to \application{Face Recognition}, we still see performance begin to degrade after 4$\times$ acceleration.  By 12$\times$, the average latency is not yet infinite, though it does exceed 3000~ms.  But at 16$\times$ and above, the average end-to-end latency again tends toward infinity.

While we see the broker waiting time begin to grow at 6$\times$ and particularly 12$\times$ acceleration (suggesting that the brokers are probably again facing a storage bottleneck), the real bottleneck in this application arises from a new category entirely.  In \Fig{fig:or_speedup_frame}, we have added a ``Delay'' category for latency components; this component represents the time between when a frame (or set of frames) was supposed to start processing in ingestion and when it actually starts processing.  This delay arises from a set of frames taking longer than 33.3~ms and delaying the start of the subsequent set of frames.

This ingestion delay represents a new manifestation of the AI tax we had not seen previously.  For every set of frames, we have opted to send each frame to the brokers separately; this ensures that they can be fully load balanced by the brokers.  However, the time required to send the full set of frames rapidly grows with the larger set sizes, so that it soon exceeds the time allotted to each set.  Kafka is well designed, however, so the producers and the brokers manage to intelligently batch the frames before sending them.  But by 16$\times$ speedup, the AI tax from sending so many items so rapidly has overwhelmed the capacity of the producers to keep up.

We see this bottleneck reflected in the throughput as well.  At 1$\times$, the throughput is 630~FPS, as expected.  That scales pretty well up to 8$\times$ speedup, but it falls short of what is expected at 12$\times$ and the system saturates by 16$\times$ speedup.

\section{AI-Centric Data Center Design}
\label{sec:solution}

\begin{figure}[t]
    \centering
    \small
    \begin{subfigure}{1.0\textwidth}
        \centering
        \captionsetup{belowskip=3pt,aboveskip=-3pt}
        \includegraphics[width=0.64\textwidth]{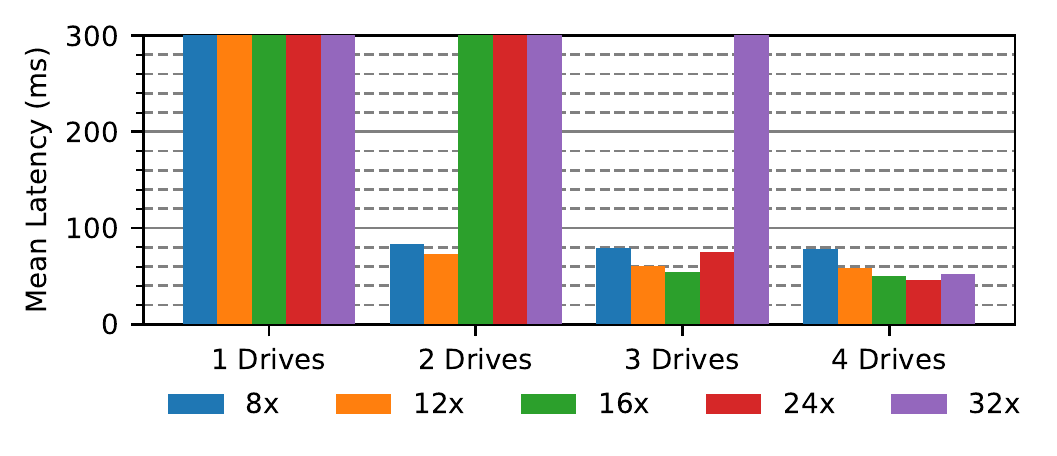}
        \caption{Using additional storage devices on each node to add bandwidth.}
        \label{fig:fr_speedup_frame_fixed:drives}
    \end{subfigure}
    \begin{subfigure}{1.0\textwidth}
        \centering
        \captionsetup{belowskip=3pt,aboveskip=-3pt}
        \includegraphics[width=0.64\textwidth]{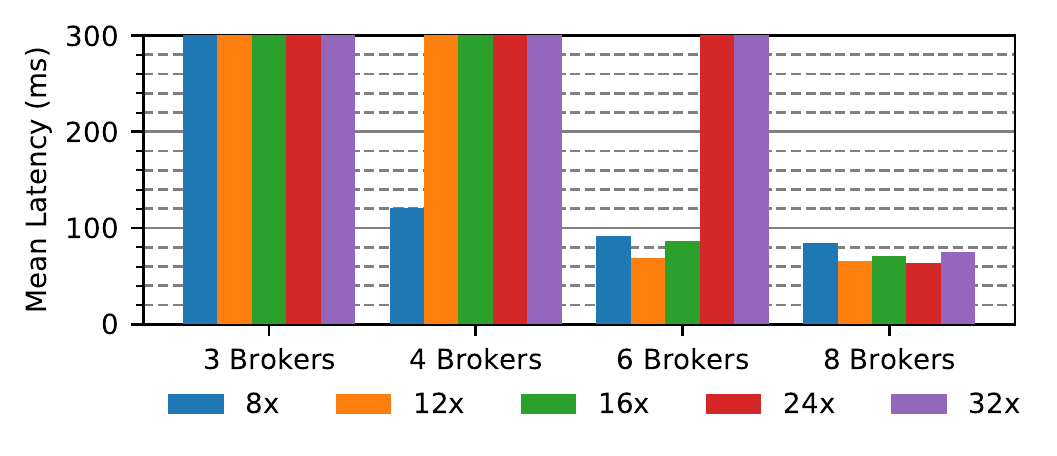}
        \caption{Using additional brokers to handle added bandwidth requirements.}
        \label{fig:fr_speedup_frame_fixed:brokers}
    \end{subfigure}
    \begin{subfigure}{1.0\textwidth}
        \centering
        \captionsetup{belowskip=3pt,aboveskip=-3pt}
        \includegraphics[width=0.64\textwidth]{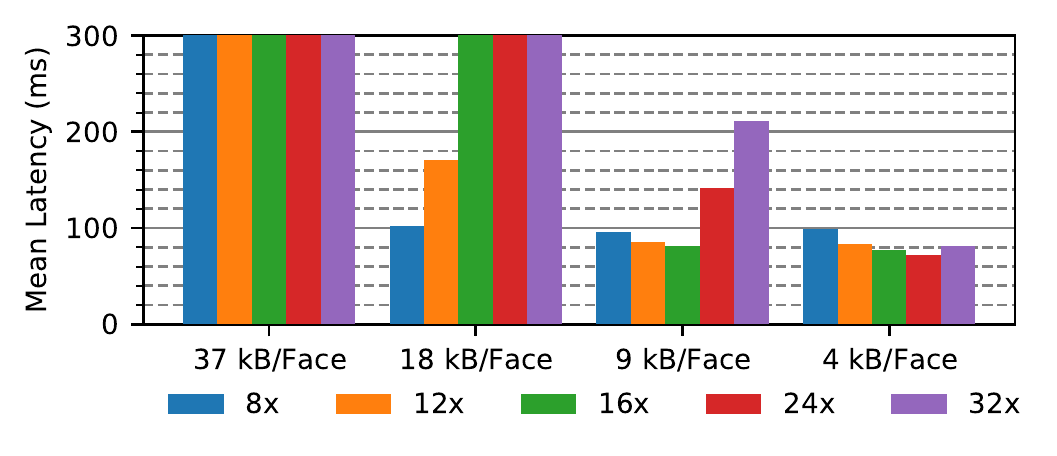}
        \caption{Decreasing thumbnail sizes to offset higher bandwidth demands.}
        \label{fig:fr_speedup_frame_fixed:sizes}
    \end{subfigure}
    \caption{Average frame latency under accelerated AI.  The higher bandwidth demands can be accommodated by allocating extra bandwidth or reducing face thumbnail sizes.
    }
    \label{fig:fr_speedup_frame_fixed}
\end{figure}

As future accelerators emerge, we seek to unlock higher speedups. In \Sec{sec:accelerate}, we saw that in an accelerated AI environment, the AI tax overwhelms the communication mechanism; in particular, the storage medium is quickly saturated at relatively modest emulated compute acceleration speeds. So in \Sec{sec:solution:unlocking} we explore two avenues to overcoming this bottleneck.  Implementing these solutions to the tax translates to actual monetary cost, as shown in \Sec{sec:solution:homogeneous}.  We show in \Sec{sec:solution:purposebuilt} that a purpose-built edge data center can address the tax with increased capacity to handle accelerated compute at lower total cost of ownership (TCO).

\subsection{Unlocking Higher Speedups}
\label{sec:solution:unlocking}

There are three ways to deal with the limitation in the storage bandwidth: (1)~utilize faster storage in the existing brokers, either through a faster storage medium (e.g.\ Intel Optane~\cite{intel_optane}) or through multiple drives operating in parallel; (2)~create more storage bandwidth by allocating additional brokers; or (3)~decrease the size of the face thumbnails, thus demanding less bandwidth.  We explore all three methods (\Fig{fig:fr_speedup_frame_fixed}), first increasing the installed drive count from one to four to provide greater bandwidth to each broker node, then increasing the broker count from three to eight across distinct broker nodes, and finally decreasing the face thumbnails down to one-eighth their original size.

\paragraph{Increasing the Bandwidth}
\label{sec:solution:bandwidth}
The effect of additional storage bandwidth on the existing nodes is captured in \Fig{fig:fr_speedup_frame_fixed:drives}.  For these experiments, we instantiated additional broker instances on each broker node (one for each drive) to ensure that each drive is given the same access to compute and memory resources; in practice, only one broker should be instantiated per node to avoid replicating data on the same node.  In the figure, we start with 8$\times$ speedup---the speedup that sent latency to infinity in the previous experiments---and increase the emulated speedup to 32$\times$.  With just one NVMe drive, the average end-to-end frame latency is infinite (depicted by the latency bar extending beyond the limits of the chart) at 8$\times$ and all higher speedups.  These experiments rely on additional drives being installed only in the brokers---in our case, there are three of them; the remainder of the servers remain unaltered from their original configuration.
In increasing the storage bandwidth by going from one drive to two drives, both 8$\times$ and 12$\times$ speedups are ``unlocked''---the system gains the ability to support compute of these speeds.  With three drives, the system supports up to 24$\times$ speedup, and with four drives, 32$\times$ is unlocked.

\paragraph{Spreading the Load}
\label{sec:solution:brokers}
Rather than installing additional drives in each of the brokers, we can instantiate additional brokers in the data center.  This spreads out the load on storage to more brokers and hence more drives.  Returning each broker to its default storage configuration (one NVMe drive), we repeat our experiments with four, six, and eight brokers.

With three brokers, as we saw before, any acceleration factor at or above 8$\times$ leads to infinite latency.  A small 33\% increase in the broker count (going from three to four brokers), however, allows the system to handle the 8$\times$ factor, while a 2$\times$ broker increase allows for up to 16$\times$ acceleration.  At eight brokers, the system can handle a 32$\times$ factor.

We find an important distinction between adding additional drives to existing brokers and adding additional brokers: the latter is more efficient.  To achieve the ability for the system to support 32$\times$ accelerated AI compute, we had to increase the number of drives by a factor of four; in contrast, we had to increase the broker count by 2.7$\times$ (going from three brokers to eight) for the same performance achievement.  The significantly lower increase in storage bandwidth in the increased-brokers approach indicates that brokers may also benefit from having additional compute capacity, memory bandwidth, or network bandwidth available.

\paragraph{Decreasing the Demand}
\label{sec:solution:sizes}

There exists one additional possibility for reining in the bandwidth demands on the storage: decrease the volume of data that needs to be stored.  Rather than spreading the data among additional brokers, the data volume can be reduced by decreasing the average size of face thumbnails.  \Fig{fig:fr_speedup_frame_fixed:sizes} shows the effect of face sizes at one-half, one-quarter, and one-eighth their original size.  Similar to increasing the bandwidth in each broker, we see that the smaller face sizes use a smaller portion of the available bandwidth and so increase the maximum supportable speedup, but without instantiating additional brokers or installing additional storage devices.

This solution, however, comes with serious trade-offs.  Decreasing face size using compression would require additional compute time, potentially offsetting much or all of the accelerator gains.  Decreasing face size by using smaller thumbnails changes the algorithm and can detrimentally impact accuracy.  Due to these severe limitations of this approach, we will focus on the previous two solutions.

\subsection{The Cost of the AI Tax in the Data Center}
\label{sec:solution:homogeneous}

A typical and simple approach that customers rely on to build an edge data center is to aim for homogeneity across servers (i.e., all of the server components are literally identical across the machines).  But in a specialized application domain, such as edge video analytics, this ignores the unique characteristics of the applications and either significantly over-provisions some resources or severely handicaps application performance, leading to suboptimal TCO.

\begin{table}[t]
    \small
    \centering
    \caption{Homogeneous data center equipment.  In a homogeneous data center similar to (but larger than) our own, there is considerable expense in ensuring that all components are equally equipped.  The equipment cost of a 1024-node data center would be around US\$30.9 million.}
    \begin{tabu}{ l r r }
        \toprule
        \textbf{Component} & \textbf{Price (US\$)} & \textbf{Quantity} \\ \midrule
        Dell PowerEdge R740xd (base server) & \$28,731 & 1024 \\
        ~~~~Intel Xeon Platinum 8176 & Included & 2 \\
        ~~~~32~GB DDR4 SDRAM & Included & 12 \\ 
        ~~~~Intel SSD DC P4510 1 TB (NVMe SSD) & \$399 & 1 \\
        ~~~~Mellanox MCX415A (100~GbE adapter) & \$660 & 1 \\ \midrule
        Mellanox MSN2700-CS2F\endgraf~~~~(100~GbE switch for fat-tree topology) & \$17,285 & 160 \\
        Mellanox MCP1600 (100~GbE cable) & \$100 & 3072 \\ \midrule
        \textbf{Total} & \textbf{\$33,577,760} & \\
        \bottomrule
    \end{tabu}
    \label{tbl:equipment_homogeneous}
\end{table}

\Tbl{tbl:equipment_homogeneous} shows the basic computing and networking equipment needed to build a homogeneous 1024-node edge data center similar in compute capabilities to our own setup.  This design gives each node comparable equipment to that used in our experiments: two 28-core processors, 384~GB of RAM, 100~Gbps interconnect, and a single NVMe drive.  The nodes are connected in a three-level fat-tree topology using 32-port Mellanox Ethernet switches.  This topology ensures full-speed non-blocking network connectivity to each node.

Using an open source TCO calculator from Coolan~\cite{coolantcomodel} to include power (servers, networking equipment, cooling, etc.), rack equipment, cabling costs, etc.\ and assuming a three-year amortization life, we estimate a yearly cost of US\$10.2 million for server equipment, US\$1.3 million for network equipment, and US\$1.4 million for power, for a total yearly cost of US\$12.9 million.

While common wisdom regarding data centers suggests that the majority of the TCO is spent on power (including powering cooling equipment), simple analysis shows this is not necessarily the case.  Each of the servers in our hypothetical data center is equipped with a 750~watt power supply, while Mellanox reports that its routers can consume a maximum of 398~watts~\cite{mellanoxspecs}.  This yields a total maximum power consumption of 921~kW. Cooling is estimated to require approximately as much power as the compute resources~\cite{hpepoweradvisor, dataspancoolingcosts}, bringing the total to 1842~kW.  Assuming US\$0.10 per kilowatt hour, operating the data center would cost US\$184 per hour or US\$1.61 million per year under maximum load.

To accommodate up to 32$\times$ accelerated compute in AI, we must either install three additional drives in each node (to maintain homogeneity) or designate a large number of the nodes as brokers.  Adding the additional NVMe drives costs US\$1.23 million.  Instead, we designate 157 of the nodes as brokers, 289 as producers, and 578 as consumers.  This maintains the ratio of each node type as in our original \application{Face Recognition} experiments (15 producer and 30 consumer nodes, though with 8 brokers instead of 3) to enable support for 32$\times$ accelerated AI.  Extrapolating from \Fig{fig:fr_speed_bw_util:network}, we estimate each producer and consumer node will consume approximately 4~Gbps of network bandwidth and each broker node 24~Gbps.  The broker nodes demand less than 9~Gbps (or 1.1~GB/s) of storage write bandwidth.

\subsection{AI-Specific Edge Data Center}
\label{sec:solution:purposebuilt}

\begin{table}[t]
    \small
    \centering
    \caption{Video analytics-targeted data center equipment.  We use network splitter cables to supply 50~Gbps network to the brokers and, in combination with slower switches, 10~Gbps network to the compute nodes.  This offers significant savings on network equipment.  Further, we only install NVMe drives in the broker nodes, which are equipped with less compute power than the compute nodes.}
    \begin{tabu}{ l r r }
        \toprule
        \textbf{Component} & \textbf{Price (US\$)} & \textbf{Quantity} \\ \midrule
        Dell PowerEdge R740xd (compute server) & \$28,731 & 867 \\
        ~~~~Intel Xeon Platinum 8176 & Included & 2 \\
        ~~~~32~GB DDR4 SDRAM & Included & 12 \\
        ~~~~Mellanox MCX411A (10~GbE adapter) & \$180 & 1 \\ \midrule
        Dell PowerEdge R740xd (broker server) & \$11,016 & 157 \\
        ~~~~Intel Xeon Bronze 3104 & Included & 2 \\
        ~~~~32~GB DDR4 SDRAM & Included & 12 \\
        ~~~~Mellanox MCX413A (50~GbE adapter) & \$395 & 1 \\
        ~~~~Intel SSD DC P4510 1 TB (NVMe SSD) & \$399 & 4 \\ \midrule
        Mellanox MSN2700-CS2F\endgraf~~~~(100~GbE switch) & \$17,285 & 28 \\
        Mellanox MSN2700-BS2F\endgraf~~~~(40~GbE switch) & \$10,635 & 14 \\ \midrule
        Mellanox MFA7A20-C010\endgraf~~~~(optical splitter 100~GbE to 2x 50~GbE) & \$1,165 & 7 \\
        Mellanox MC2609130-003\endgraf~~~~(copper splitter 40~GbE to 4x 10~GbE) & \$90 & 217 \\
        Mellanox MCP7H00-G002R\endgraf~~~~(copper splitter 100~GbE to 2x 50~GbE) & \$140 & 79 \\
        Mellanox MFA1A00-C030\endgraf~~~~(optical 100~GbE interconnect) & \$515 & 192 \\
        \midrule
        \textbf{Total} & \textbf{\$27,878,431} & \\
        \bottomrule
    \end{tabu}
    \label{tbl:equipment_ai}
\end{table}

\begin{figure*}
    \small
    \centering
    \includegraphics[width=1.0\textwidth]{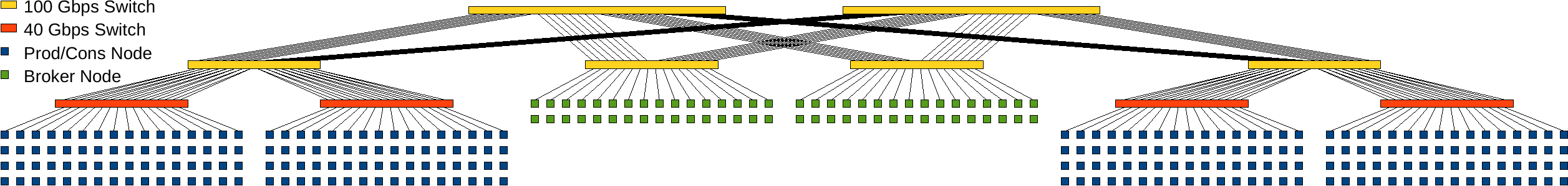}
    \caption{Possible network configuration for an accelerated AI-centric edge data center.  This network configuration is fundamentally a fat-tree built from 100~Gbps switches.  However, since no node needs the full 100~Gbps bandwidth, we subdivide it.  Mellanox offers splitter cables that split, for example, a 100~Gbps connection into two 50~Gbps or four 25~Gbps connections.  Each pair of broker nodes shares a 100~Gbps port.  Rather than providing each producer/consumer node with 50 or even 25~Gbps, we further subdivide the network connection speed using 32-port 40~Gbps switches.
    To provide full bandwidth to the 40~Gbps switches, each 100~Gbps port is split into two 50~Gbps connections, both of which are connected to the same 40~Gbps switch.
    Thus, the aggregate bandwidth provided to a 40~Gbps switch is 800~Gbps, of which the switch can use 640~Gbps.  A 100~Gbps switch can connect two 40~Gbps switches.  The slower switches use four-way splitter cables providing each producer/consumer node with 10~Gbps network.}
    \label{fig:onprem_network}
\end{figure*}

The homogeneous data center was designed to be generic, capable of executing a variety of application classes; hence, we had to adapt the application to the data center, resulting in hugely over-provisioned network and storage.  The producers and consumers constitute over 84\% of the data center and use only 4\% of the available network capacity and essentially none of the storage bandwidth.  The brokers use a respectable 24\% of the network capacity and basically all of the storage bandwidth but use very little of the compute capacity.  This shows extremely inefficient allocation of limited resources in the data center.  But we can do better.

Instead of forcing the application to fit into an existing data center, we propose building a data center that fits the application.  We recommend a \emph{purpose-built data center} that specifically targets the broker-specific AI tax (the demand for storage bandwidth).  The AI tax, if not accounted for, can translate to non-trivial real-world costs that can drastically affect end users' needs.  In contrast, by understanding the AI tax and designing to it, we demonstrate that a moderately-sized edge data center can be purpose-built to better address the AI tax while yielding meaningful cost savings.

In \Tbl{tbl:equipment_ai} we see the equipment needed for this setup, designed to support up to 32$\times$ AI acceleration.  In this scenario, we utilize the same highly parallel servers as in \Tbl{tbl:equipment_homogeneous} for producers and consumers but limit the network bandwidth on these nodes to only 10~Gbps and install only basic storage for operating each server.  The broker nodes, in contrast, are built on far less parallel but still impressive CPUs, while enjoying 50~Gbps network connections and four NVMe SSDs.

We illustrate in \Fig{fig:onprem_network} a simple network solution that could provide the designated bandwidths to each server.  At its heart, the network is still a fat-tree built from 100~Gbps Mellanox switches, but, using Mellanox splitter cables and slower 40~Gbps switches, broker nodes are provided with 50~Gbps connections while producer and consumer nodes get 10~Gbps connections.  A single edge switch can connect 32 broker nodes or 128 producer/consumer nodes.  We can thus build the complete data center using a two-level fat-tree of just 28 100~Gbps switches (12 edge and 16 core); seven edge switches connect to a total of fourteen 40~Gbps switches and five connect to the 157 brokers.

In designing this purpose-built data center, we wanted to avoid limiting potential advancements or upgrades during the lifetime of the data center.  We designed it with double the anticipated requirements for network and storage bandwidth.
The brokers were designed to accommodate the 32$\times$ speedup in two separate ways.  First, we maintained the higher ratio of brokers to compute nodes, just as we did in the homogeneous design; second, we allocated four times the storage devices and bandwidth to each broker.  Either one of these solutions on its own would have been adequate to accommodate the compute speedup.  Furthermore, by giving 50~Gbps and 10~Gbps network connections to the broker and compute nodes, respectively, we have allowed them to grow to double their anticipated needs.  In combination, we have given the data center the ability to adapt to unanticipated application speedups during its intended lifetime.

Our purpose-built data center incurs an equipment cost of US\$27.9 million with a yearly power cost of US\$1.4 million for a three-year amortized yearly total cost of ownership of US\$10.8 million.  This is 16.6\% lower than the TCO of the homogeneous data center while being better equipped to handle future accelerated compute.

\section{Related Work}
\label{sec:related}

We build on prior work that enabled and rapidly expanded AI and ML applications.  Unlike most of the prior work, however, we explore the implications of accelerating AI computation and how it affects an end-to-end application flow.  We present related work in five categories: (1)~AI and ML benchmarking, (2)~integrating AI and ML, (3)~end-to-end application flow studies, (4)~exploiting heterogeneity, and (5)~edge data centers.

\paragraph{Benchmarking}
\label{sec:related:benchmarking}
MLPerf is one of the leading resources for benchmarking ML-related compute~\cite{janapareddi2019mlperf, mlperfinference}.   It provides flexibility for benchmarking a variety of hardware across a variety of ML kernels, but it entirely ignores the issue of end-to-end application behavior and performance.  In our work we demonstrate the central importance of understanding the end-to-end application, showing that each ML kernel can constitute a relatively small portion of the pipeline and that truly optimizing ML performance requires a more holistic view of the system.

\paragraph{Integrated AI}
\label{sec:related:ai}
In presenting the scale and deployment of ML workloads at Facebook, Hazelwood et al.\ acknowledged the importance of pre-processing data for training and emphasized its stress on storage, network, and CPU~\cite{hazelwood2018applied}.  They acknowledged the potentially high latency of inference for top quality models but did not expose the overhead of pre- and post-processing.  Nor did they discuss the resource requirements of streaming inference workloads.  We emphasize both of these to show how they can pose a barrier to overall performance improvement from AI acceleration.

Microsoft recognizes the importance of latency in the data center particularly as it applies to deep neural networks~\cite{brainwave}.  Chung et al.\ presented Microsoft's Project Brainwave which implements DNNs largely in FPGAs distributed throughout the data center, emphasizing the importance of accelerating increasingly complex DNNs~\cite{chung2018serving}.
In contrast, this work emphasizes the importance of the enabling code for AI and assumes that accelerating AI and ML will be successful, instead looking at its ultimate impact on the larger workflow.

\paragraph{End-to-End Application Flows}
\label{sec:related:endtoend}
Though not specific to AI workloads, Kanev et al.\ offered a comprehensive look at the trends of warehouse-scale computing at Google~\cite{kanev2015profiling}.  They quantified data center ``taxes''---overheads that come with applications but do not directly contribute to the end result, including compression, communication serialization, and remote procedure calls.  We show that the brokers act as a tax, coordinating the activities of a distributed application.

Other work has broken down the end-to-end latency of requests, at various levels of granularity, ranging from evaluation of Internet speed and programming language patterns to operating system scheduling and memory latency~\cite{chow2014mystery, li2014tales}.  This more closely matches our contribution, though our analysis is restricted to latency within the data center and is focused specifically on the common communication base (Ap\-a\-che Kafka) of open source streaming frameworks in an effort to bring perspective to end-to-end AI application flows.

\paragraph{Heterogeneous Execution}
Prior works sought to exploit the heterogeneity in a data center, producing benefits in speed, energy consumption, and operating costs~\cite{mars2011heterogeneity, haque2017exploiting}.  Where these papers sought to capitalize on unintentional heterogeneity (arising from workload co-location and, for example, later upgrades), we extol the benefits of intentionally designing an on-premise data center with heterogeneous servers and network.  We thereby add hardware cost savings to the existing benefits of data center heterogeneity.

\paragraph{Edge Data Centers}
\label{sec:related:onprem}
Hewlett Packard Enterprise recently demonstrated that cloud-based computing is often not the most cost-effective solution~\cite{hpe2018onprem}.  Their analysis showed for a well-utilized edge data center, TCO can be drastically lower than comparable capabilities in the cloud.  Our work extends that idea, showing how the on-premise data center can be specifically tailored to the needs of AI applications.

\section{Conclusion}
\label{sec:conclusion}

It is easy to get caught up in the excitement of AI and ML; this work has brought context to those advancements, elucidating an AI tax, and serves as a call to action to address limiters of performance in realistic, edge data center deployments of AI applications.  Streaming AI applications are only possible with the support of pre- and post-processing code, which is far from trivial in both latency and compute cycles and relies almost exclusively on the CPU for all of the processing.  AI applications will likely be composed of multiple inference stages, each with its own characteristics and overheads.  And the enabling substrate for managing AI applications in a data center sees hot spots in both network and storage that could soon become bottlenecks if not addressed.

\bibliographystyle{ACM-Reference-Format}
\bibliography{tex/references}

\end{document}